\documentstyle[12pt,bezier]{article}
\def \thesection {\arabic{section}.}
\def \thesubsection {\thesection\arabic{subsection}.}

\def \sect #1 {\setcounter{equation} 0\section{#1}}
\def \appendix #1#2 {\par\bigskip\bigskip\noindent
                    {\Large {\bf Appendix {#1}. {#2} }}
                    \def\thesubsection{}
                    \def\theequation{{#1}.\arabic{equation}}
                    \setcounter{equation} 0 \par\bigskip\noindent}
\def \figures {\begin{center}{\Large {\bf Figures: }}\end{center}\par\bigskip}
\def \theequation {\thesection\arabic{equation}}   
\def \be  {\begin{equation}}
\def \ee  {\end{equation}}
\def \ba  {\begin{eqnarray}}
\def \ea  {\end{eqnarray}}
\def \baa {\begin{eqnarray*}}
\def \eaa {\end{eqnarray*}}
\def \bb  {\begin{thebibliography}}
\def \eb  {\end{thebibliography}}
\newcommand \ci [1] {\cite{#1}}
\newcommand \bi [1] {\bibitem{#1}}
\def \lab #1 {\label{#1}}
\newcommand\re[1]{(\ref{#1})}
\def \qqquad {\qquad\quad}

\newcommand\lr[1]{{\left({#1}\right)}}

\def \tr {\mbox{tr\,}}
\def \Im {\mbox{Im\,}}
\def \Re {\mbox{Re\,}}
\def \vev #1 {\langle{#1}\rangle}
\def \VEV #1 {\left\langle{#1}\right\rangle}
\newcommand \ket[1] {|{#1}\rangle}

\def \e {\mbox{e}}
\def \CO {{\cal O}}

\def \CD {{\cal D}}

\newcommand\sign{\mbox{sign}\,}
\newcommand \eps \varepsilon
\def \fracs #1#2 {\mbox{\small $\frac{#1}{#2}$}}
\newcommand\partder[1]{{\partial\over\partial{#1}}}
\def \bin #1#2 {{\left({#1}\atop{#2}\right)}}
\def \as {\relax\ifmmode\alpha_s\else{$\alpha_s${ }}\fi}
\def \alpi {\frac \as \pi}
\def \al #1 {\frac {\as({#1})}{\pi} }
\def \ds #1 {\ooalign{$\hfil/\hfil$\crcr$#1$}}
\def \MS {\overline{\rm MS}}
\def \QCD {\mbox{{\tiny QCD}}}

\catcode`\@=11\relax
\renewcommand\eqnarray{\stepcounter{equation}\let\@currentlabel=\theequation
              \global\@eqnswtrue
              \global\@eqcnt\z@\tabskip\@centering\let\\=\@eqncr
              $$\halign to \displaywidth\bgroup\@eqnsel\hskip\@centering
              $\displaystyle\tabskip\z@{##}$&\global\@eqcnt\@ne
              \hskip 0.7 \arraycolsep \hfil${##}$\hfil&\global\@eqcnt\tw@
              \hskip 0.7 \arraycolsep $\displaystyle\tabskip\z@{##}$\hfil
              \tabskip\@centering&\llap{##}\tabskip\z@\cr}

\begin{document}

\def\thefootnote{\fnsymbol{footnote}}
\thispagestyle{empty}
\hfill\parbox{35mm}{{\sc ITP--SB--94--42}\par
                         hep-ph/9409446  \par
                         September, 1994}
\vspace*{40mm}
\begin{center}
{\LARGE High energy scattering in QCD and \\[5mm]
cross singularities of Wilson loops}
\par\vspace*{20mm}\par
{\large I.~A.~Korchemskaya}
\footnote{On leave from the Moscow Energetic Institute,
          Moscow, Russia}
\par\bigskip\par\medskip
{\em Dipartimento di Fisica, Universit\`a di Parma and \par
INFN, Gruppo Collegato di Parma, I--43100 Parma, Italy}
\par\bigskip\par
and
\par\bigskip\par
{\large G.~P.~Korchemsky}
\footnote{On leave from the Laboratory of Theoretical Physics,
          JINR, Dubna, Russia}
\par\bigskip\par\medskip
{\em Institute for Theoretical Physics, \par
State University of New York at Stony Brook, \par
Stony Brook, New York 11794 -- 3840, U.S.A.}
\end{center}
\vspace*{20mm}

\begin{abstract}
We consider elastic quark-quark scattering at high energy and fixed
transferred momentum. Performing factorization of soft gluon exchanges
into Wilson line expectation value we find that there is one-to-one
correspondence between high energy asymptotics in QCD and renormalization
properties of the so called cross singularities of Wilson lines. Using
this relation we show that the asymptotic behavior of the quark-quark
scattering amplitude is controlled by a $2\times 2$ matrix of the cross
anomalous dimensions. We evaluate the matrix of cross anomalous dimension
to two-loop order and study the properties of the obtained expressions
to higher loop order.
\end{abstract}

\newpage

\def\thefootnote{\arabic{footnote}}
\setcounter{footnote} 0

\section{Introduction}

Recent experimental data from Tevatron and HERA renewed our
interest to theoretical interpretation of the well known phenomena in the
physics of strong interactions like growth of the total hadronic
cross section at high energy and increasing of the structure function
of deep inelastic scattering at small values of the Bjorken variable $x$.
In both cases we deal with the asymptotic behavior of hadronic
scattering amplitudes at high energy $s$ and fixed transferred momenta $t$
the so called Regge behavior in high-energy QCD \ci{Regge}. This problem is one
of the longstanding in QCD and previous attempts to solve it have led
to the discovery of the BFKL pomeron \ci{BFKL}. However, found in the leading
logarithmic approximation, $\as \ll 1$ and $\as\log s\sim 1$,
the BFKL
pomeron leads to the expressions for the scattering amplitudes which
violate the unitarity Froissart bound. To preserve the unitarity of
the $S-$matrix, one has to go beyond the leading logarithmic approximation
calculating the scattering amplitudes. The problem turns out to be very
complicated and the following approaches have been proposed:
\begin{itemize}
\item
   Generalized leading logarithmic approximation \ci{GLLA,Bar},
   in which we start with the
   leading logarithmic result and add ``minimal'' number of nonleading
   corrections in order to unitarize the result;
\item
   Two-dimensional reggeon field theory \ci{two-dim},
   in which the high-energy scattering
   of gluons is described by the $S-$matrix of the effective two-dimensional
   field theory corresponding to dynamics of transverse gluonic degrees of
   freedom;
\item
   Quasielastic unitarity approximation \ci{qua}, in which one takes into
   account only contributions of soft gluons with transverse momenta
   much smaller hadronic scale.
\end{itemize}
These schemes are approximate and they were developed to resum special
class of corrections which are expected to determine the high-energy
asymptotics of the scattering amplitudes. In each case, QCD is replaced
by simplified effective theory which inherits a rich structure
of the original theory.

Recently \ci{Lip,FK}, a progress has been made using the
generalized leading logarithmic approximation. It was found that in
this approximation high energy QCD turns out to be a completely
integrable model equivalent to the one-dimensional Heisenberg magnet of spin
$s=0$. This opens the possibility to apply powefull quantum inverse
scattering method and calculate the hadronic scattering amplitudes by
means of the Generalized Bethe Ansatz \ci{FK}.

There were different proposals for the effective two-dimensional
reggeon field theory \ci{two-dim,Ver} and the hope is that the model
can be solved exactly.  Although to the lowest order of PT the
effective reggeon theory reproduces the QCD expressions for the
scattering amplitudes, the all order perturbative solution is not
available yet.

The present paper is devoted to the calculation of the hadronic
scattering amplitudes in the quasielastic unitarity approximation
\ci{qua}. In this approximation, we use the parton model and
approximate the hadron-hadron interaction as the scattering of
partons. Partons interact with each other by exchanging soft gluons
with energy which is much smaller than parton energy, $s$, but much
bigger than the QCD scale $\Lambda_{\rm QCD}$ in order for the
perturbative QCD to be applicable. Being calculated to the lowest
orders of PT, the parton-parton scattering amplitudes get large
perturbative corrections $\lr{\as \log s \log t}^n$ due to soft gluon
exchanges \ci{qua} and these Sudakov logarithms need to be resummed to
all orders of PT. The resummation of leading logarithmic corrections
can be performed using the ``evolution equation'' technique
\ci{qua,evol} but we don't have a regular way to resum and control
nonleading logarithmic corrections to the scattering amplitude. A new
approach which enables us to take into account systematically both
leading and nonleading logarithmic corrections to the parton-parton
scattering amplitudes has been proposed in \ci{K}. In this approach,
the partonic scattering in QCD is described by the dynamics of the
gauge invariant collective variables defined as
\be
W=W(C_1,C_2,\ldots,C_n)\equiv \vev{0|T\
\tr_{R_1}P\e^{ig\oint_{C_1}dx\cdot A(x)} \cdots
\tr_{R_n}P\e^{ig\oint_{C_n}dx\cdot A(x)} |0} \, ,
\lab{def}
\ee
the so called {\it Wilson loop functions\/}. Here, the path-ordered
exponentials are evaluated along the closed paths $C_1$ $\ldots$ $C_n$
in Minkowski space-time and the gauge fields are defined in the
irreducible representations $R_1$ $\ldots$ $R_n$ of the $SU(N)$ gauge
group. One should stress that it is not for the first time when one
dealt with this object. There were attempts \ci{Mig} in 80's to study
the nonperturbative dynamics of QCD in terms of Wilson loops. The
interest to the Wilson loops was renewed after it was recognized
\ci{pQCD} that the same loop functions enter into consideration in
perturbative QCD when we study the effects of soft gluons in hardonic
processes. Interacting with soft gluons described by gauge field
$A_\mu(x)$, partons behave as relativistic charged classical
particles and the path-ordered exponentials appear as eikonal phases of
partons. The integration paths $C_1$ $\ldots$ $C_n$ have a meaning of
semi-classical trajectories of partons and the choice of the
representations $R_1$ $\ldots$ $R_n$ is fixed by their charges. In
particular, $R$ is fundamental and adjoint representation of the
$SU(N)$ group for quarks and gluons, respectively.

The paper is organized as follows. In sect.2 we consider the amplitude
of near forward elastic quark-quark scattering and show its relation with
the renormalization properties of the so called cross singularities of
Wilson loops. In sects.3 and 4 we analyze the cross singularities of
Wilson loops to the lowest orders of PT and find two-loop expression for
the matrix of cross anomalous dimensions. The properties of the obtained
expressions which can be generalized to all orders of PT are studied in
sect.5. The details of the calculations of the two-loop Feynman diagrams
contributing to the Wilson lines are given in Appendices A and B.
Large $N$ limit of the Wilson line is analysed in Appendix C.

\section{High energy scattering in QCD}

Let us briefly summarize the results of the paper \ci{K} concerning
the relation between high energy behavior of the scattering amplitudes
in QCD and the renormalization properties of the Wilson loops.  We
consider the near forward elastic quark-quark scattering in the
following kinematics:
$$
s,\ m^2 \gg -t \gg \lambda^2 \gg \Lambda^2_{\QCD}.
$$
Here, $s=(p_1+p_2)^2$ is the invariant energy of quarks with mass $m$,
$t=(p_1-p_1')^2$ is the transferred momentum and $\lambda^2$ is the IR
cutoff (characteristic hadronic scale)
which is introduced to regularize infrared divergences.
Notice that we don't need to fix the relation between $s$ and $m$ and
quarks may be heavy. The
scattering amplitude $T_{ij}^{i'j'}$ contains the color indices of the
incoming and outgoing quarks. The quarks interact with each other by
exchanging soft gluons with the total momentum $q=p_1-p_1'$. In the center
of mass frame of incoming quarks, for $s\gg -t$, the transferred momentum
has the light-cone components $q=(q^+,q^-,\vec q)$ with
$q^\pm=\frac1{\sqrt 2}(q^0\pm q^3)=0$
and
$q^2=-\vec q^2=t$. Since the transferred momentum is much smaller than
the energies of incoming quarks, the only effect of their interaction with
soft gluons is the appearance of additional phase in the wave functions of
the incoming quarks. This phase, the so called eikonal phase, is equal to
a Wilson line $P\e^{i\int_C dx\cdot A(x)}$ defined in the fundamental
representation of the $SU(N)$ group and evaluated along the classical
trajectory $C$ of the quark. We combine the eikonal phases of both quarks
and obtain the representation for the quark-quark scattering amplitude as
\ci{K}
\be
T_{ij}^{i'j'} \lr{\frac{s}{m^2},\frac{q^2}{\lambda^2}}
=\sinh\gamma\int d^2z\, \e^{-i\vec z\cdot\vec q}\
W_{ij}^{i'j'}(\gamma,\vec z^2\lambda^2)  \, ,
\qquad t=-{\vec q\,}^2\,,
\lab{main}
\ee
where the line function $W_{ij}^{i'j'}$ is given by
\be
W_{ij}^{i'j'}(\gamma,\vec z^2\lambda^2)=
\langle 0| T
\left[P\ \e^{ig\int_{-\infty}^\infty d\alpha\,
v_1\cdot A(v_1\alpha)}\right]_{i'i}
\left[P\ \e^{ig\int_{-\infty}^\infty d\beta\,
v_2\cdot A(v_2\beta+z)}\right]_{j'j}
|0\rangle\,.
\lab{def1}
\ee
Here, two Wilson lines are evaluated along infinite lines
in the direction of the quark velocities $v_1=p_1/m$ and $v_2=p_2/m$
and the integration paths are separated by the impact vector
$z=(0^+,0^-,\vec z)$ in the transverse direction,
$v_1\cdot z=v_2\cdot z=0$.

Being expanded in powers of the gauge field, expression \re{main}
reproduces the quark-quark scattering amplitude in the eikonal
approximation. Integration over two-dimensional impact vector $z$ was
introduced in \re{main} to ensure the total transferred momentum in
the $t-$channel to be $q=(0^+,0^-,\vec q)$. Indeed, it follows from
\re{main} that the total momentum of gluons $k_{\rm tot.}$ propagating in
the $t-$channel is restricted by the following $\delta-$functions:
$$
\delta(k_{\rm tot.}\cdot v_1)
\delta(k_{\rm tot.}\cdot v_2)
\delta(\vec k_{\rm tot.}-\vec q)
=\frac{1}{\sinh \gamma}
\delta(k_{\rm tot.}^+)\delta(k_{\rm tot.}^-)\delta(\vec k_{\rm tot.}-\vec q)
=\frac{1}{\sinh \gamma}\delta^{(4)}(k_{\rm tot.}-q)\,,
$$
where the first $\delta-$function comes from integration over $\alpha$ in
\re{def1}, the second one from integration over $\beta$ and the last one from
integration over $\vec z$ in \re{main}. To compensate the additional factor,
$\sinh \gamma$, in the r.h.s. of this relation, the same factor was
introduced in the expression \re{main}.

The scattering amplitude \re{main} depends on the quark velocities $v_1$
and $v_2$, the transferred momentum $\vec q$ and IR cutoff $\lambda$. These
variables give rise to only two scalar dimensionless invariants:
$(v_1 v_2)=s/m^2$ and $t/\lambda^2$ as explicitly indicated in \re{main}.
The $s-$dependence of the amplitude comes through the dependence
on the angle $\gamma$ between quark $4-$velocities $v_1$ and $v_2$ defined
in Minkowski space-time as
\be
(v_1\cdot v_2)=\cosh\gamma\,.
\lab{angle}
\ee
For arbitrary velocities $v_1$ and $v_2$, the angle $\gamma$ may take
real and complex values and in the limit
of high energy quark-quark scattering ($s\gg m^2$) we have
\be
\gamma = \log\frac{s}{m^2} \gg 1\,.
\lab{gl}
\ee
To find the asymptotic behavior of the scattering amplitude \re{main}
one needs to know the vacuum expectation value of the Wilson lines
\re{def1}.
Let us consider first a special case of the elastic electron-electron
scattering in QED.

\subsection{High-energy scattering in QED}

One may use the expression \re{main} for the scattering
amplitude in this case, but the calculation of the Wilson line
\re{def1} is simplified
because for abelian gauge group a path-ordered exponent coincides with an
ordinary exponent. Moreover, since photons don't interact with each other,
the calculation of Wilson lines in QED is reduced to the integration of a
free photon propagator $D_{\mu\nu}(x)$
along the path and the result of calculation is
\be
W_{\rm QED}(\gamma,z^2\lambda^2)=
\exp\lr{\frac{(ie)^2}2 \int dx^\mu\int dy^\nu D_{\mu\nu}(x-y)}
=\lr{\lambda^2 {\vec z\,}^2/4} {}^{i\frac{e^2}{4\pi}\coth\gamma}\,,
\lab{W-QED}
\ee
where $\lambda$ is a fictitious photon mass. After substitution of this
relation into \re{main} and integration over impact vector we get
the expression for the scattering amplitude in QED \ci{QED}
\be
T_{\rm QED}\lr{\frac{s}{m^2},\frac{t}{\lambda^2}} = ie^2\frac{\cosh\gamma}{t}
\left(\frac{-t}{\lambda^2}\right)^{-i\frac{e^2}{4\pi}\coth\gamma}
\frac{\Gamma(1+i\frac{e^2}{4\pi}\coth\gamma)}
     {\Gamma(1-i\frac{e^2}{4\pi}\coth\gamma)}\,.
\lab{A-QED}
\ee
In the limit $s/m^2\to\infty$ or, equivalently $\gamma\to\infty$, the
expectation value \re{W-QED} does not depend on $s$ and, as a consequence,
the scattering amplitude has the asymptotics $T/T_{\rm Born}\sim s^0$ with
$T_{\rm Born}=\frac{ie^2}{t}\frac{s}{m^2}$ the amplitude in the Born
approximation. This implies that photon is not reggeized in QED \ci{QED}.

\subsection{Wilson lines in perturbative QCD}

The evaluation of the Wilson line in perturbative QCD is much more
complicated than in QED. To explain the approach \ci{K} for the
evaluation of the Wilson line \re{def1}, let us consider as an
example the one-loop calculation of $W$ in the Feynman gauge.
Using the definition \re{def1}, we get
\be
W_{\rm 1-loop}=I\otimes I+(t^a\otimes t^a)\frac{g^2}{4\pi^{D/2}}\Gamma(D/2-1)
\lambda^{4-D}
\int_{-\infty}^{\infty}d\alpha\int_{-\infty}^{\infty}d\beta
\frac{(v_1v_2)}{[-(v_1\alpha-v_2\beta)^2+{\vec z\,}^2+i0]^{D/2-1}}\,,
\lab{W-one}
\ee
where a direct product of the gauge generators defined in the fundamental
representation takes care of the color indices of the quarks. In this
expression, gluon is attached to both Wilson lines at points $v_1\alpha$
and $v_2\beta+z$ and we integrate the gluon propagator
$v_1^\mu v_2^\nu D_{\mu\nu}(v_1\alpha-v_2\beta-z)$
over positions of these points.
To regularize IR divergences we introduced the dimensional regularization
with $D=4-2\varepsilon,$ ($\varepsilon < 0$) and $\lambda$ being the IR
renormalization parameter. There is another contribution to $W_{\rm 1-loop}$
corresponding to the case when gluon is attached by both ends to one of
the Wilson lines. A careful treatment shows that this contribution
vanishes. Performing integration over parameters $\alpha$ and $\beta$ in
\re{W-one} we get
\be
W_{\rm 1-loop}(\gamma,\lambda^2{\vec z\,}^2)
=I\otimes I+(t^a\otimes t^a)\alpi (-i\pi\coth\gamma)
\Gamma(-\varepsilon)
(\pi\lambda^2{\vec z\,}^2)^{\varepsilon}\,.
\lab{W-z}
\ee
The integral over $\alpha$ and $\beta$ in \re{W-one} has an infrared
divergence coming from large $\alpha$ and $\beta$. In the dimensional
regularization, this divergence appears in $W_{\rm 1-loop}$ as a pole
in $(D-4)$
with the renormalization parameter $\lambda$ having a sense of an IR cutoff.
After renormalizaion in the ${\MS}-$scheme, this result coincides up to
color factor with the lowest term in the expansion of the analogous
expression \re{W-QED} in QED in powers of $e^2$.

The evaluation of the Wilson line \re{def1} in QCD is based on
the following observation. The one-loop expression \re{W-z}
for the line function is divergent for $\vec z=0$ and $\varepsilon<0$.
This divergence has an ultraviolet origin
because it comes from integration over small $\alpha$ and $\beta$ in
\re{W-one}, that is from gluons propagating at short distances
($v_1\alpha-v_2\beta$) between quark trajectories as $\alpha, \ \beta\to 0$.
One should notice that
the ultraviolet (UV) divergences of Wilson lines at $z=0$ have nothing
to do with the ``conventional'' ultraviolet singularities in QCD. For
$z=0$, the integration paths of Wilson lines in \re{def1} cross each other
at point $0$ and UV divergences, the so called cross divergences,
appear when one integrates gluon propagators along integration path at
the vicinity of the cross point \ci{Br}.
We note that the cross divergences do not appear in $W_{\rm 1-loop}$
for nonzero $z$ because, as it enters into the integral \re{W-one},
nonzero ${\vec z\,}^2$ regularizes gluon propagator at short distances
as $\alpha$, $\beta\to 0$.

Thus, ${\vec z\,}^2$ has a meaning of an UV cutoff for the one-loop line
function $W$ and the dependence of the original line function
$W(\gamma,\lambda^2{\vec z\,}^2)$ on $z^2$ is in
one-to-one correspondence with the dependence of the same Wilson line but
with $z=0$ on the UV cutoff which we have to introduce to regularize the
cross singularities. This property suggests the following way for calculation
of the $z-$dependence of the Wilson lines \re{def1}.
First, we put $z=0$ in the definition \re{def1} of the line function and
introduce the regularization of
cross singularities of $W$. Then, we renormalize cross singularities and
identify UV cutoff with impact vector as $\mu^2=1/{\vec z\,}^2$.
We conclude that the $z-$dependence of the line function $W$ and, as
consequence, the $t-$asymptotics of the scattering amplitude \re{main}
are controlled
by the renormalization properties of the cross singularities of Wilson loops.

\subsection{Renormalization properties of Wilson lines}

The renormalization properties of the loop functions in QCD have been
studied in detail \ci{Br} and can be summarized as follows.  Being
expanded in powers of gauge potential, $W(C_1,\ldots,C_n)$ can be
expressed as multiple integrals of the Green functions along the paths
$C_1$ $\ldots$ $C_n$.  The loop function is not a well defined object
because it contains different kinds of divergences. First of all, it
has the conventional ultraviolet (UV) divergences of the Green
functions in perturbative QCD. Moreover, as it was first found in
\ci{Pol}, the Wilson loops have very specific UV divergences which
depend on the properties of the integration paths $C$ in the
definition \re{def1}.  For instance, the expansion of the loop
function to the lowest order of PT involves the contour integral
$\oint_C dx_\mu \oint_C dy_\nu\
D^{\mu\nu}(x-y)$ which is potentially divergent at $x=y$ due to
singular behavior of the gluon propagator at short distances.  The
divergences come either when $y$ approaches $x$ along the path $C$ or
when the path crosses itself and both $x$ and $y$ approach the cross
point. In both cases divergences have an UV origin because they come
from integration at short distances.  In the first case, UV
divergences can be absorbed into renormalization of the coupling
constant and gauge fields provided that the integration path is smooth
everywhere. Otherwise, if the path is smooth everywhere except of a
one point where it has a cusp, additional UV divergences, the
so-called ``cusp'' singularities, appear which can be renormalized
multiplicatively \ci{multi}.  The corresponding anomalous dimension,
$\Gamma_{\rm cusp}(\gamma,g)$, the cusp anomalous dimension is a gauge
invariant function of the coupling constant and the cusp angle. If the
integration path crosses itself (or the paths $C_1$ $\ldots$ $C_n$
cross each other), additional ``crossed'' singularities appear in the
loop function \ci{Br}.

It was found \ci{pQCD} that these renormalization properties of the
Wilson loops which look like their undesired features are of the most
importance in perturbative QCD. The same function, the cusp anomalous
dimension of a Wilson line, appears when one studies the infrared (IR)
singularities of on-shell form factors or the velocity-dependent
anomalous dimension in the effective heavy quark field theory. The
anomalous dimensions of the Wilson lines govern the IR logarithms in
the same manner as the anomalous dimensions of the local composite
twist-2 operators control the collinear singularities one encounters
studying the structure functions of deep inelastic
scattering. Moreover, there is a remarkable relation \ci{K} between
the high-energy behavior of the scattering amplitudes in the
quasielastic unitarity approximation and the renormalization
properties of the ``cross'' divergences of the loop function $W$.

\subsection{Cross anomalous dimension}

The renormalization properties of the cross divergences
were studied in \ci{Br}. It was shown that the loop function \re{def}
with all Wilson loops defined in the {\it fundamental\/} representation of
the $SU(N)$ group ($R_1=\cdots=R_n=F$) and paths having a cross point, is
mixed under renormalization with loop functions evaluated along
the integration paths which are different from original paths at the
vicinity of the cross point. As an important example we consider
the Wilson lines $(W_1)^{i'j'}_{ij}$ and $(W_2)^{i'j'}_{ij}$ defined
in fig.1. Although the integration paths in fig.1 have an infinite length
and they are open we could get two Wilson loops by projecting the line
functions $(W_1)^{i'j'}_{ij}$ and $(W_2)^{i'j'}_{ij}$
onto the color structure $\delta_{i'j}\delta_{j'i}$.

We choose the Wilson lines in fig.1(a) and 1(b) to be defined in the
fundamental representation of the $SU(N)$ gauge group. According to the
general analysis \ci{Br}, the line function of fig.1(a) has cross
singularities and
under their renormalization it is mixed with the Wilson line of fig.1(b).
As a consequence, the renormalized line functions $W_1$ and $W_2$ satisfy
the following renormalization group (RG) equation \ci{Br}
\be
\lr{ \mu\partder{\mu}+\beta(g)\partder{g} }W_a
=-\Gamma_{\rm cross}^{ab}(\gamma,g)W_b , \qquad
a,\, b=1,\ 2\,,
\lab{RG}
\ee
where $W_a\equiv \lr{W_a}^{i'j'}_{ij}$ and $\mu$ is renormalization point.
Here, $\Gamma_{\rm cross}(\gamma,g)$ is
the cross anomalous dimension which is a gauge invariant $2\times2$ matrix
depending only on the coupling constant and the angle $\gamma$ between lines
at the cross point.
It is important to notice that the RG equation \re{RG} is valid only in the
fundamental
representation of the gauge group and its generalization to another
representations is uknown.

\subsection{Evaluation of the scattering amplitude}

We recall that the evaluation of the scattering amplitude \re{main}
is based on the
following property of the line function $W$. Using the definition \re{def1}
of $W$ we introduce into consideration the Wilson line $W_1$ in fig.1(a).
The only
difference between them is that the integration paths in \re{def1} do not
cross each other for nonzero $z$. For $z\neq 0$ the Wilson line
$W(\gamma,\vec z^2\lambda^2)$ is UV finite but for $z=0$ it coincides
with $W_1$ and has cross divergences. The distance between quark trajectories,
$\vec z^2$, in the definition \re{def1} of $W$ plays a role of UV cutoff for
cross singularities and the $z-$dependence of $W$ is the same as the
dependence of the Wilson line
$W_1(\gamma,\mu^2/\lambda^2)$ on the renormalization
parameter $\mu$. Solving the RG equation \re{RG} for
$W_1(\gamma,\mu^2/\lambda^2)$ with the boundary conditions,
$W_1(\gamma,1)=\delta_{i'i}\delta_{j'j}$ and
$W_2(\gamma,1)=\delta_{i'j}\delta_{j'i}$,
we substitute $\mu^2=1/\vec z^2$ and find
the following expression for the scattering amplitude \ci{K}
\be
T_{ij}^{i'j'} \lr{\frac{s}{m^2},\frac{\vec q^2}{\lambda^2}}
=\sinh\gamma\int d^2\vec z\, \e^{-i\vec z\vec q}\
\lr{ A_{11}(\gamma,\vec z^2\lambda^2)\ \delta_{ii'}\delta_{jj'}
+A_{12}(\gamma,\vec z^2\lambda^2)\ \delta_{ij'}\delta_{ji'}}
\,,
\lab{res0}
\ee
where $A_{11}$ and $A_{12}$ are elements of the $2\times 2$ matrix
\be
A_{\rm qq}\lr{\gamma,z^2\lambda^2}=T\ \exp\lr{-\int_\lambda^{1/z}
\frac{d\tau}{\tau}\Gamma_{\rm cross}(\gamma,g(\tau))}.
\lab{A}
\ee
Notice, that the matrices $\Gamma_{\rm cross}(\gamma,g(\tau))$ do not commute
with each other for different values of $\tau$ and this makes difficult
to evaluate the time-ordered exponent in \re{A}. The expression \re{res0} for
the quark-quark scattering amplitude can be easily generalized to describe
the quark-antiquark scattering. We notice that antiquark with a 4-velocity
$v_2$ can be treated as a quark moving backward in time with velocity $-v_2$.
As a result, to get the quark-antiquark scattering amplitude one has
to replace one of the quark velocities $v_2\to-v_2$ in expressions \re{res0}
and \re{A}.
Since the scattering amplitude \re{res0} depends on the angle $\gamma$
between velocities, this transformation corresponds to the replacement
$\gamma\to i\pi-\gamma$ in the expression \re{A} for the matrix $A$,
or equivalently in cross anomalous dimension:
$$
A_{\rm q\bar q}\lr{\gamma,z^2\lambda^2}=T\ \exp\lr{-\int_\lambda^{1/z}
\frac{d\tau}{\tau}\Gamma_{\rm cross}(i\pi-\gamma,g(\tau))}\,,
$$
where $\gamma$ is the angle between velocities of incoming quark and
antiquark.

The expression \re{res0} takes into account all $\log t$ and $\log s$
corrections to the scattering amplitude. We conclude from \re{A} that
it is the matrix of the cross anomalous dimensions which governs the high
energy behavior of the scattering amplitude. As it follows
from \re{res0}, the $s-$dependence of the amplitude comes from the
$\gamma-$ dependence of $\Gamma_{\rm cross}(\gamma,g)$ while the
$t-$dependence originates from the evolution of the coupling constant.
To find the asymptotic behavior of the scattering amplitude \re{res0} we need
to know the large $\gamma-$behavior of $\Gamma_{\rm cross}(\gamma,g)$ to all
orders of perturbation theory. To this end, we perform the one-loop calculation
of $\Gamma_{\rm cross}(\gamma,g)$ in the next section and try to find the
properties of this matrix which can be generalized to higher loop orders.

\section{One-loop calculation of the cross anomalous dimension}

To find the cross anomalous dimension we evaluate the line functions of
fig.1(a) and 1(b) to the lowest order of PT and substitute them into
\re{RG}. In the Born approximation the line functions have trivial expressions
\be
W_1^{(0)}=\delta_{i'i}\delta_{j'j}\equiv \ket{1},\qquad
W_2^{(0)}=\delta_{i'j}\delta_{j'i}\equiv \ket{2}\,,
\lab{Born}
\ee
where $\ket{1}$ and $\ket{2}$ denote two states in color space.
To one-loop order the line functions are given by path integrals similar to
that in \re{W-one}. To evaluate the integrals we introduce the following
parameterization of the integration paths of fig.1,
\be
C_{\rm I}=v_1 \alpha\,,      \qquad
C_{\rm II}=v_2 \beta\,,  \qquad
C_{\rm III}=v_1 \beta\,,  \qquad
C_{\rm IV}=v_2 \alpha\,,
\lab{para}
\ee
where $0<\alpha<\infty$ and $-\infty<\beta<0$ and where $v_1$ and $v_2$ are
quark velocities.

The integration paths in fig.1(a) and 1(b) cross each other at point $0$ and
go to infinity along the vectors $v_1$ and $v_2$. As a consequence, the line
functions $W_1$ and $W_2$ have UV divergences due to the cross point and IR
divergences due to the infinite length of the paths.
To regularize the cross singularities we use the dimensional
regularization with $D=4-2\varepsilon$ and in order to regularize IR
divergences we
introduce a fictitious gluon mass $\lambda$. We fix the Feynman gauge
and define the gluon propagator in the coordinate representation as
$
D_{\mu\nu}(x)=g_{\mu\nu}D(x)
$
with
\be
D(x)=-i\int\frac{d^Dk}{(2\pi)^D}\frac{\e^{-ikx}}{k^2-\lambda^2+i0}
    =-\frac1{4\pi^{D/2}}\int_0^{-i\infty} d\alpha\ \alpha^{D/2-2}
     \exp\lr{x^2\alpha-\frac{\lambda^2}{4\alpha}}\,.
\lab{prop}
\ee
After the decomposition of the integration path of fig.1(a) as
$C=C_{\rm I}+C_{\rm II}+C_{\rm III}+C_{\rm IV}$
we get the set of Feynman diagrams of fig.2 which contribute to one-loop
corrections to the line function $W_1$.
Then, the contribution of the diagram of fig.2(1) to $W_1$ is given by
$$
(ig)^2t^a_{i'i}t^a_{j'j}
\int_{C_{\rm I}}dx_\mu\int_{C_{\rm II}}dy_\nu\ D^{\mu\nu}(x-y)\equiv
-g^2(v_1v_2)t^a\otimes t^a\int_0^\infty d\alpha
\int_{-\infty}^0 d\beta\ D(v_1\alpha-v_2\beta)\,,
$$
where integration is performed over positions of gluons on the paths
$C_{\rm I}$ and $C_{\rm II}$ and the direct product of the gauge group
generators takes care of color indices of the line function.
Substituting the gluon propagator \re{prop} and integrating over
$\alpha$ and $\beta$ we get the expression for the ``Feynman integral''
\be
F_1=(ig)^2\int_{C_{\rm I}}dx_\mu\int_{C_{\rm II}}dy_\nu\ D^{\mu\nu}(x-y)
 =-\alpi
\lr{\frac{4\pi\mu^2}{\lambda^2}}^\eps
\frac{\Gamma(1+\eps)}{2\eps} \gamma\coth\gamma\,.
\lab{F1}
\ee
Here, the cross singularity comes from the integration over small $x-y$ and it
appears in the final expression as a pole in $\varepsilon$.
Calculation of the remaining diagrams of fig.2 is analogous and gives
the following results
\be
F_2=-\alpi\lr{\frac{4\pi\mu^2}{\lambda^2}}^\eps
\frac{\Gamma(1+\eps)}{2\eps}(i\pi-\gamma)\coth\gamma
, \qquad
F_3=-2\,F_4=-\alpi\lr{\frac{4\pi\mu^2}{\lambda^2}}^\eps
\frac{\Gamma(1+\eps)}{2\eps}\,.
\lab{F2-F4}
\ee
We notice that there are simple relations between these expressions:
$F_2(\gamma)=-F_1(i\pi-\gamma)$, $F_3=F_1(\gamma=0)$. In Appendix A
we explain their origin and show that analogous relations exist
between two-loop Feynman integrals.

To evaluate the one-loop correction to the line function $W_1$ we take the
expressions \re{F1} and \re{F2-F4} and sum them with appropriate color and
combinatoric factors:
$$
W_1^{(1)}=t^a_{i'i}t^a_{j'j}(2 F_1 + 2 F_2)+
C_F\ \delta_{i'i}\delta_{j'j} (2 F_3 + 4 F_4)\,,
$$
where $C_F=t^a t^a$ is the quadratic Casimir operator in the fundamental
representation of the $SU(N)$ gauge group. This expression contains a pole
in $\varepsilon$ which is subtracted in
the $\MS-$scheme. Finally, the renormalized one-loop expression for the line
function $W_1$ is given by
\be
W_1^{(1)}=-\alpi\log\frac{\mu^2}{\lambda^2}(t^a_{ii'}t^a_{jj'})\
i\pi\coth\gamma\,.
\lab{W1(1)}
\ee
The calculation of the line function $W_2$ involves the integration over
the path
of fig.1(b) which is different from the path of fig.1(a) only at the cross
point. The contribution of the individual Feynman diagram of fig.2 to the
line function $W_1$ has the form of the product, $W= F\cdot C\cdot w$,
of the corresponding Feynman integral
($F$), the color factor ($C$) and the combinatorial weight of the diagram
($w$).
The difference between the paths of fig.1(a) and 1(b) at the cross point is not
important when one calculates the Feynman integral $F$ and the combinatorial
weight $w$ but it does affect the color factors of the corresponding
diagrams. This means, that to any order of perturbation theory the same
Feynman integrals contribute to both line functions and the only difference
between $W_1$ and $W_2$ is that integrals enter with different
color factors. In particular, using the expressions \re{F1} and \re{F2-F4} we
get the one-loop expression for the line function $W_2$ as
$$
W_2^{(1)}=C_F\ \delta_{i'j}\delta_{j'i}(2 F_1 + 4 F_4)+
          t^a_{i'j}t^a_{j'i}(2 F_2 + 2 F_3)\,.
$$
After substitution of the Feynman integrals \re{F1} and \re{F2-F4} into this
expression we subtract a pole in the $\MS$ scheme and obtain the renormalized
line function
\be
W_2^{(1)}=\alpi\log\frac{\mu^2}{\lambda^2}
\left[
-C_F\ \delta_{i'j}\delta_{j'i}\,(\gamma\coth\gamma-1)
+t^a_{i'j}t^a_{j'i}(\gamma\coth\gamma-i\pi\coth\gamma-1)
\right]\,.
\lab{W2(1)}
\ee
The obtained one-loop expressions \re{W1(1)} and \re{W2(1)}
for the line functions are valid in
an arbitrary representation of the $SU(N)$ group
after proper redefinition of the Casimir operator $C_F$. The reason why the
fundamental representation plays a special role comes from the fact that
the direct product of the gauge group generators can be decomposed into
the sum of invariant tensors
\be
t^a_{ij} t^a_{kl} = -\frac1{2N}\delta_{ij}\delta_{kl}
                      +\frac12\delta_{il}\delta_{jk}\,.
\lab{dec}
\ee
After substitution of this decomposition into \re{W1(1)} and \re{W2(1)} we
find that one-loop correction to $W_1$ contains a color structure
of $W_2$ in the Born approximation, eq.\re{Born}, and vice versa.
As a consequence,
both line functions are mixed with each other under renormalization.
In an arbitrary representation of the gauge group we do not have a relation
similar to \re{dec} and the renormalization properties of the line functions
become complicated.

Substituting the one-loop results, \re{W1(1)} and \re{W2(1)},
for the line functions into the RG
equation \re{RG} we find the one-loop expression for the matrix of cross
anomalous dimension:
\ba
&&\Gamma_{\rm cross}(\gamma,g)=\alpi \Gamma(\gamma)\,,
\lab{cross(1)}
\\
&&\Gamma(\gamma)=
\lr{ \begin{array}{cc} -\frac{i\pi}{N}\coth\gamma   &    i\pi \coth\gamma
\\             -\gamma\coth\gamma+1+i\pi\coth\gamma & N(\gamma\coth\gamma-1)
-\frac{i\pi}{N}\coth\gamma \end{array} }\,.
\nonumber
\ea
Here, $\gamma$ is the angle between quark velocities $v_1$ and $v_2$
in Minkowski space-time defined in \re{angle}.

\subsection{Properties of the one-loop $\Gamma_{cross}$}

The obtained expression \re{cross(1)}
for one-loop $\Gamma_{\rm cross}$ obeys the following
interesting properties. All elements of the matrix $\Gamma_{\rm cross}$ have
nontrivial imaginary parts which change their sign under the replacement
$\gamma\to i\pi-\gamma$:
$$
\Im \Gamma_{\rm cross}(i\pi-\gamma,g)=-\Im \Gamma_{\rm cross}(\gamma,g)\,.
$$
As was shown in sect.2.5, this transformation of the angle allows us to
relate quark-quark and quark-antiquark scattering amplitudes.
Calculating the determinant of the matrix \re{cross(1)}
we obtain the expression
\be
\det\Gamma_{\rm cross}(\gamma,g)=\lr{\alpi}^2\pi^2\coth^2\gamma\
\lr{1-\frac1{N^2}}\,,
\lab{det}
\ee
which, first, has zero imaginary part and, second, is invariant under
$\gamma\to i\pi-\gamma$. This property implies that the elements of the
matrix \re{cross(1)} are ``fine tunned''. An additional
argument in favour of this observation comes from the consideration
of the large $\gamma$ behavior of $\Gamma_{\rm cross}$.
In the limit of large $\gamma$ different elements of the matrix
have the following behavior
$$
\Gamma^{11}_{\rm cross}\sim\Gamma^{12}_{\rm cross}\sim \CO(\gamma^0), \qqquad
\Gamma^{12}_{\rm cross}\sim\Gamma^{22}_{\rm cross}\sim \CO(\gamma)\,,
$$
which leads to the asymptotics $\det\Gamma_{\rm cross}\sim\CO(\gamma)$
in general. However, the one-loop result \re{det}
implies that the leading term $\CO(\gamma)$
vanishes and the determinant has the asymptotics
$\det\Gamma_{\rm cross}\sim\CO(\gamma^0)$ while
$\tr\Gamma_{\rm cross}\sim\CO(\gamma)$. The eigenvalues of the matrix
$\Gamma(\gamma)$ satisfy the characteristic equation
\be
\Gamma_\pm^2
-\Gamma_\pm \lr{ N\gamma\coth\gamma-N-\frac{2i\pi}{N}\coth\gamma }
+\lr{\pi\coth\gamma}^2=0\,.
\lab{ein}
\ee
Solving this equation we find that in the large $\gamma$
limit both eigenvalues
have positive real part (Re $\Gamma_\pm >0$). Moreover, one of the
eigenvalues of matrix $\Gamma$ is much larger than the second one:
\ba
\Gamma_+ &=&
N\gamma - \lr{N+\frac{2i\pi}{N}}
-\pi^2\frac{N^2-1}{N^3}\gamma^{-1}
+ \CO\lr{\gamma^{-2}},
\nonumber
\\
\Gamma_- &=&
\pi^2\frac{N^2-1}{N^3}\gamma^{-1}
+ \CO\lr{\gamma^{-2}}\,,
\lab{ein-l}
\ea
with $\gamma=\log(s/m^2)$ in the limit $s\gg m^2$.

\subsection{Asymptotic behavior of the scattering amplitudes}

Let us substitute the one-loop expression \re{cross(1)} for the matrix
$\Gamma_{\rm cross}$ into \re{res0} and \re{A}.
One-loop matrices $\Gamma_{\rm cross}(\gamma,g(\tau))$ commute with each
other for different $\tau$ and we can omit $T-$ordering in the
definition \re{A} of the matrix $A$. After diagonalization of the one-loop
matrix $\Gamma(\gamma)$ by a proper unitary transformation we get
$$
A_{\rm qq}\lr{\gamma,{\lambda^2}{z^2}}=
\frac{\Gamma_+-\Gamma(\gamma)}{\Gamma_+-\Gamma_-}\
\exp\lr{ -\Gamma_- \int_\lambda^{1/z}\frac{d\tau}{\tau}\al{\tau} }
+
\frac{\Gamma_--\Gamma(\gamma)}{\Gamma_--\Gamma_+}\
\exp\lr{ -\Gamma_+ \int_\lambda^{1/z}\frac{d\tau}{\tau}\al{\tau} }
\,,
$$
where the eigenvalues $\Gamma_\pm$ were defined in \re{ein}.
Finally, we find the following expression for
the scattering amplitude \re{res0}
\be
T_{ij}^{i'j'}=\delta_{i'i}\delta_{j'j}\ T^{(0)}
             +t^a_{i'i} t^a_{j'j}\ T^{(8)}
\,,
\lab{res1}
\ee
where two invariant amplitudes correspond to the exchange in the
$t-$channel by states with the quantum numbers of vacuum and gluon
and are given by
\be
T^{(0)}=-\frac{\sinh\gamma}{t}\frac{\Gamma_+\Gamma_-}{\Gamma_+-\Gamma_-}
(T_+-T_-),
\qquad
T^{(8)}=2i\pi\frac{\cosh\gamma}{t}\frac{\Gamma_+T_+-\Gamma_-T_-}
{\Gamma_+-\Gamma_-}\,.
\lab{res2}
\ee
Here the notation was introduced for the Fourier transforms
\be
T_\pm=\frac{t}{\Gamma_\pm}\int d^2z\ \e^{-i\vec z\cdot\vec q}
\exp\lr{ -\Gamma_\pm\int_\lambda^{1/z}\frac{d\tau}{\tau} \al{\tau} }\,.
\lab{rr}
\ee
Using one-loop expression for the coupling constant we get
\be
T_\pm=\frac{t}{\Gamma_\pm}\int d^2z\ \e^{-i\vec z\cdot\vec q}
\left( \frac{\log 1/z^2\Lambda_{\QCD}^2}{\log\lambda^2/\Lambda_{\QCD}^2}
\right)^{-2\Gamma_\pm/\beta_0}\,.
\lab{tro}
\ee
Let us first consider the properties of the obtained expressions \re{res2}
and \re{rr} in the leading $\log t$ approximation,
$\as\log t/\lambda^2 \sim 1$. In this limit we may freeze the
argument of the coupling constant in \re{rr} and
perform Fourier transformation to get
\be
T_{\pm} = 2\as
\exp\lr{-\frac{\as}{2\pi}\,\Gamma_\pm\lr{\frac{s}{m^2}}
\,\log\frac{-t}{\lambda^2}}\
\frac{\Gamma\lr{1+\frac{\as}{2\pi}\Gamma_\pm}}
     {\Gamma\lr{1-\frac{\as}{2\pi}\Gamma_\pm}}
\,.
\lab{lla}
\ee
This expression is similar to that \re{A-QED} in QED and the only difference
is that in QCD the $s-$dependence of the amplitudes is governed by the
asymptotic behavior of the eigenvalues $\Gamma_\pm$.
In the case of abelian gauge group the Wilson lines of figs.1(a) and 1(b)
coincide and the cross anomalous dimension in QED is a c-number and not
a matrix.
As a result,
the scattering amplitudes have different high energy behavior in QCD and QED.

Expanding \re{lla} in powers
of $\as$ we find that to the lowest order of PT
$$
T_\pm=2\as + \CO(\as^2)\,,
$$
where higher order corrections involve $\log s$ and $\log t$ terms.
After substitution of these values into \re{res2} we recover
the expressions for the scattering amplitude in the Born approximation
$$
T^{(0)}=0 + \CO(\as^2)\,,\qqquad
T^{(8)}=\frac{ig^2}{t}\frac{s}{m^2} + \CO(\as^2)\,.
$$
To find the behavior of the scattering amplitude \re{res2} in the limit of
high energies $s \gg m^2$, or equivalently for large cross angle \re{gl},
we start with the leading $\log t$ and $\log s$ approximation,
$\as\log s/m^2 \log t/\lambda^2 \sim 1$. We use \re{ein-l} and substitute
$\Gamma_+=N\log s/m^2$ and $\Gamma_-=0$ into \re{lla} and \re{res2}.
The result has the standard reggeized form \ci{PT}:
\be
T^{(0)}_{\rm LL}=0\,,\qquad
T^{(8)}_{\rm LL}=T_{\rm Born}\cdot  \lr{\frac{s}{m^2}}^{\alpha(t)}
\,,
\lab{LL}
\ee
with the Regge trajectory $\alpha(t)
=-\frac{\alpha_s}{2\pi}N\log\frac{-t}{\lambda^2}$.
We can improve the leading logarithm approximation \re{LL} by
taking into account nonleading
corrections to \re{lla} in the limit
$
\as \log s/m^2 \sim \as \log t/\lambda^2 \sim 1\,.
$
Using the large energy behavior of the eigenvalues
\re{ein-l} we find the following expressions for the amplitudes of singlet
and octet exchanges
\be
T^{(0)}=-T_0\
\Gamma_-\lr{T_+ - T_-}\, ,
\qquad
T^{(8)}=2i\pi T_0\
\lr{T_+-\frac{\Gamma_-}{\Gamma_+}T_-} \,,
\lab{NL}
\ee
with $T_0=\frac{s}{2 m^2 t}$ and $\Gamma_\pm$ defined in \re{ein-l}.
The functions $T_+$ and $T_-$ have a Regge like behavior \re{lla}
and for $\Gamma_+ \gg \Gamma_-$ we have $T_+\ll T_-$. For $\Gamma_-=0$
the expression \re{NL} coincides with the leading log result \re{LL}.
We stress that the amplitude $T^{(0)}$ of the singlet exchange
is unvanishing only due to nonleading logarithmic corrections.
Moreover, as follows from \re{NL}, the amplitude $T^{(0)}$ gets dominant
contribution from $T_-$ and not from $T_+$. The same is true for
the amplitude $T^{(8)}$ of the octet exchange in which exponential fall
of $T_+$ can not compete  with the factor $\Gamma_-/\Gamma_+$ in
front of $T_-$. Thus, the high-energy asymptotic behavior of the scattering
amplitude \re{NL} is dominated by the contribution of $T_-$ and
as a consequence the amplitude of the octet exchange
is suppressed by the factor $1/\Gamma_+=\CO(\log^{-1} s/m^2)$ as compared
to that to the amplitude of the singlet exchange. Comparing this behavior
with \re{LL} we conclude that nonleading logarithmic corrections
drastically change the leading log asymptotics of the scattering amplitude.

Expression \re{lla} was found under condition that we neglect the
$\tau-$dependence of the coupling constant $\as(\tau)$ in \re{rr}.
We notice, however, that \re{lla} becomes divergent for the special value
of the coupling constant $\as=2\pi/\Gamma_\pm$. To understand the
origin of this divergence we deduce from \re{rr} that it is the
distance between quark trajectories, $z^2$, which fixes the scale of
the coupling constant. Then, for the perturbative expansion to be meaninfull
this distance should be much bigger than the QCD scale $1/\Lambda_{\rm QCD}^2$.
On the other hand, to get the scattering amplitude in \re{rr} we have to
integrate in \re{rr} over all possible $z^2$ including large distances.
For the $z-$integral in \re{rr} to get dominant contribution in PT one has
to impose additional conditions on $\Gamma_\pm$ and $\vec q^2$,
or equivalently on $s$ and $t$. Otherwise, the perturbative expression
\re{rr} become divergent at large distances. The analysis of PT applicability
conditions for \re{rr} can be performed similar to the analysis given in
\ci{DY} for the Drell-Yan process.

The properties of the scattering amplitude are closely related to
the particular form of the one-loop cross anomalous dimension \re{cross(1)}.
Natural question arises whether the asymptotic behavior of the scattering
amplitudes will be changed after we take into account higher order corrections
to the cross anomalous dimension. In the next section we answer this
question by performing the calculation of the matrix of cross anomalous
dimension to two-loop order in the Feynman gauge.

\section{Two-loop calculation of the cross anomalous dimension}

To find two-loop expression for the cross anomalous dimension we have to
calculate the line functions $W_1$ and $W_2$ to the second order of PT
by taking into account the next terms in the expansion of Wilson lines in
powers of the coupling
constant and effects of gluon self-interaction. The resulting Feynman
diagrams contributing to $W_1$ are shown in figs.3--5 and we divide them into
three different groups
\begin{enumerate}
\item
   QED-like diagrams of fig.3 in which gluons don't interact with each other;
\item
   Self-energy diagrams of fig.4 containing self-energy corrections to the
   gluon propagator;
\item
   Three-gluon diagrams of fig.5 having three-gluon vertex of self-interaction.
\end{enumerate}
Evaluating the diagrams of fig.5 containing three gluon vertex, we notice
that the Lorentz indices of the three gluon vertex
$\Gamma^{\mu\nu\rho}$ are saturated by the vectors $v_1$ or $v_2$
depending to which line the gluons are attached. As a consequence,
the diagrams with all three gluons attached to the same line vanish
as $\Gamma_{\mu\nu\rho}v_1^\mu v_1^\nu v_1^\rho=0$. We didn't
include these diagrams in fig.5.

There many different Feynman integrals corresponding to the diagrams of
figs.3--5 but, as we show in Appendix A, their calculation is reduced to
the evaluation of only few ``basic'' diagrams after we take into account the
properties of the Feynman integrals. The one-loop diagram of fig.2(1) and
two-loop diagrams of fig.3(1)--(5), fig.4(1) and fig.5(1) form the
set of the basic diagrams. Moreover, among them there are the diagrams
which have been already calculated in \ci{cusp}.

\subsection{Cusp anomalous dimension}

As was shown in \ci{cusp}, the diagrams of fig.3(3)--(5),
fig.4(1) and fig.5(1) contribute to
the two-loop cusp anomalous dimension $\Gamma_{\rm cusp}(\gamma,g)$ of
Wilson loops. 
This anomalous dimension appears in the RG equation for the Wilson
loop if the integration path has a cusp with the angle $\gamma$.

The fact that
the same diagrams of fig.3(3)--(5), fig.4(1) and fig.5(1) contribute to
the ``cusp'' and ``cross'' anomalous dimensions suggests
that there might exist a relation between these two completely
different functions of the angle $\gamma$. Indeed, the explicit
form of this relation will be suggested in sect.5. Both anomalous
dimensions depend on the
representation of the $SU(N)$ in which the Wilson loop is defined.
One might expect that if the relation between them does exist it contains
$\Gamma_{\rm cross}$ and $\Gamma_{\rm cusp}$ taken in the same representation.
In reality, as we will show in sect.5, the two-loop cross anomalous dimension
calculated in the fundamental representation is related to the two-loop cusp
anomalous dimension defined in the adjoint representation of the $SU(N)$.

The two-loop expression for $\Gamma_{\rm cusp}(\gamma,g)$ in the adjoint
representation of the $SU(N)$ is given by \ci{cusp}
\be
\Gamma_{\rm cusp}(\gamma,g)=
\alpi N(\gamma\coth\gamma-1)+
\lr{\alpi N}^2
\Phi_{\rm cusp}(\gamma)\,,
\lab{cusp1}
\ee
where $\Phi_{\rm cusp}(\gamma)
=\Phi_{\rm QED}+\Phi_{\rm self}+\Phi_{\rm 3-gluon}$
and different kinds of the diagrams (QED-like of fig.3(3)--(5),
self-energy of fig.4(1) and three-gluon of fig.5(1)) contribute to
the corresponding functions:
\ba
&&\Phi_{\rm QED}(\gamma)=\coth^2\gamma
\int_0^\gamma d\psi\,\psi(\gamma-\psi)\coth\psi
- \coth\gamma\int_0^\gamma d\psi\psi\coth\psi
+\gamma\coth\gamma-\fracs12\,,
\nonumber
\\
&&\Phi_{\rm self}(\gamma)=\frac{31}{36}(\gamma\coth\gamma-1)\,,
\lab{cusp2}
\\
&&\Phi_{\rm 3gluon}(\gamma)=-\frac12\sinh(2\gamma)
\int_0^\gamma d\psi\frac{\psi\coth\psi-1}{\sinh^2\gamma-\sinh^2\psi}
-\frac{\pi^2}{24}(\gamma\coth\gamma-1)\,.
\nonumber
\ea
The properties of $\Gamma_{\rm cusp}(\gamma,g)$ to higher orders of PT
have been studied in \ci{cusp}.

\subsection{QED like diagrams}

The same QED-like diagrams of fig.3 contribute to the line functions $W_1$
and $W_2$ with different color and combinatorial factors. The explicit
expressions for the color factors and the combinatorial weights are given
in Appendix B.

To get the contribution of the QED-like diagrams of fig.3
to the line functions $W_1$ and $W_2$
we evaluate the corresponding Feynman integrals using the relations found in
Appendix A, multiply them by color factors and combinatorial weights
and sum the resulting expressions. Finally, after lengthy calculation
we get the following
expression for the contribution of the diagrams of fig.3 to $W_1$
\ba
W_1^{\rm QED}=\lr{\alpi}^2\log^2\frac{\mu^2}{\lambda^2}
&& \left[
\lr{-\frac{N^2+1}{8N^2}\pi^2\coth^2\gamma
     -\frac{i\pi}{8}\gamma\coth^2\gamma
     +\frac{i\pi}{4}\coth\gamma}\ket{1} 
\right.
\nonumber
\\
&+&\left.\lr{\frac{\pi^2}{4N}\coth^2\gamma
    +\frac{i\pi}{8}N\gamma\coth^2\gamma
     -\frac{i\pi}{4}N\coth\gamma}\ket{2} 
\right]\,.
\lab{W1(qed)}
\ea
We notice that this expression does not contain a single logarithm of $\mu$.
Although some individual diagrams do contain $\log(\mu^2/\lambda^2)-$terms
they cancel in the total sum of the diagrams. In terms of the RG equation
\re{RG},
this means, that the QED-like diagrams don't contribute to the
two-loop expressions for the elements $\Gamma_{11}$ and $\Gamma_{12}$ of
the matrix of cross anomalous dimensions.
The contribution of the QED-like diagrams to the line function $W_2$ can
be found in analogous way. The resulting expression has the following
structure:
\be
W_2^{\rm QED}=\lr{\alpi}^2 \lr{A_{\rm QED} \log^2\frac{\mu^2}{\lambda^2}
+B_{\rm QED} \log\frac{\mu^2}{\lambda^2}}\,,
\lab{W2(qed)}
\ee
where in contrast with $W_1^{\rm QED}$ the coefficient $B_{\rm QED}$ in
front of a single logarithm of $\mu$ is different
from zero. The coefficient $A_{\rm QED}$ is given by
\baa
A_{\rm QED} &=&
\left\{
 \left(
       \frac{\pi^2}{4N}+\frac{N^2+2}{8N}i\pi\gamma-\frac18 N\gamma^2
 \right)\coth^2\gamma
+\left(
       -\frac{N^2+1}{4N}i\pi+\frac38 N\gamma
 \right)\coth\gamma
-\frac14 N
\right\}\ket{1} 
\\
&+&
\left\{
 \left(
       -\frac{N^2+1}{8N^2}\pi^2-\frac38 i\pi\gamma+\frac{1}{8}N^2\gamma^2
 \right)\coth^2\gamma
+\left(
       -\frac{3}{8}N^2\gamma+\frac{i}{2}\pi
 \right)\coth\gamma
+\frac{1}{4} N^2
\right\}
\ket{2} 
\,.
\eaa
The expression for the coefficient $B_{\rm QED}$ is complicated and it
contains the integrals $I_1$, $I_2$ and $I_3$ defined in \re{I1-I3}. We
simplify this expression using the identity \re{prope}
and after some algebra we recover the following relation between
the coefficient $B_{\rm QED}$ and the function $\Phi_{\rm QED}$
contributing to the two-loop cusp anomalous dimension \re{cusp2}:
\be
B_{\rm QED}=\frac12 N\Phi_{\rm QED}(\gamma)
   \left[\ket{1}-N\ket{2}\right]
 -\lr{\Phi_{\rm QED}(\gamma)-\Phi_{\rm QED}(i\pi-\gamma)}
   \left[N\ket{1}-\ket{2}\right]
\,.
\lab{B-qed}
\ee
After substitution of the Wilson line \re{W2(qed)} into the RG
equation \re{RG} we find that the coefficient $B_{\rm QED}$
contributes to the elements $\Gamma_{21}$ and $\Gamma_{22}$ of the
matrix of the cross anomalous dimension. The identity \re{B-qed}
implies that the contribution of the QED-like diagrams to two
different anomalous dimensions, $\Gamma_{\rm cross}$ and
$\Gamma_{\rm cusp}$, are closely related to each other.

\subsection{Self-energy diagrams}

Calculation of the self-energy diagrams of fig.4 to the Wilson lines
can be easily done using the relations found in Appendix A.
The total contribution of the diagrams of fig.4
to the line function $W_1$ is given by
\be
W_1^{\rm self}=\lr{\alpi}^2 i\pi\coth\gamma
\left(\frac5{48}\log^2\frac{\mu^2}{\lambda^2}
     +\frac{31}{72}\log\frac{\mu^2}{\lambda^2}
\right)
\left(\ket{1} - N \ket{2}\right)
\lab{W1(self)}
\ee
and it has the same color structure and the $\gamma-$dependence as the
one-loop expression \re{W1(1)} for $W_1$.
For the total contribution of the self-energy diagrams to the line function
$W_2$ we get
\ba
W_2^{\rm self}&=&\lr{\alpi}^2 \lr{\frac{5}{48}\log^2\frac{\mu^2}{\lambda^2}
                           +\frac{31}{72}\log\frac{\mu^2}{\lambda^2}}
\nonumber
\\
&\times&
\left[
N\lr{\gamma\coth\gamma-1-i\pi\coth\gamma}\ket{1}
-\lr{N^2(\gamma\coth\gamma-1)-i\pi\coth\gamma}\ket{2}
\right]\,.
\lab{W2(self)}
\ea
Having the relation \re{B-qed}, it is tempting to find the analogous relation
between the coefficient $B_{\rm self}$ in front of $\log(\mu^2/\lambda^2)$ in
$W_2^{\rm self}$ and the function $\Phi_{\rm self}$ defined in \re{cusp2}. It
can be easily checked that the relation does exist and has the following form:
\be
B_{\rm self}=\frac12 N\Phi_{\rm self}(\gamma)
   \left[\ket{1}-N\ket{2}\right]
 -\frac12\lr{\Phi_{\rm self}(\gamma)-\Phi_{\rm self}(i\pi-\gamma)}
   \left[N\ket{1}-\ket{2}\right]\,.
\lab{B-self}
\ee
However, in contrast with $B_{\rm QED}$, the second term in $B_{\rm self}$ has
additional factor one-half.

\subsection{Three-gluon diagrams}

The diagrams of fig.5 contribute to the Wilson line $W_1$ with the same
color factor defined in Appendix B.
Then the identity \re{3g} implies that the
total contribution of the diagrams with three gluon vertex to the line
function $W_1$ vanishes as
\be
W_1^{\rm 3gluon}\propto F_1^{\rm 3gluon}+F_2^{\rm 3gluon}+F_3^{\rm 3gluon}=0
\,.
\lab{W1(3g)}
\ee
For the line function $W_2$  the color factors of the diagrams of fig.5
are different and their total contribution is unvanishing
\be
W_2^{\rm 3gluon}=\lr{\alpi}^2\log\frac{\mu^2}{\lambda^2}\ B_{\rm 3gluon}\,,
\lab{W2(3g)}
\ee
where the coefficient $B_{3gluon}$ is given by
\be
B_{\rm 3gluon}=
 \frac12 N\Phi_{\rm 3gluon}(\gamma)
   \left[\ket{1}-N\ket{2}\right]
 -\lr{\Phi_{\rm 3gluon}(\gamma)-\Phi_{\rm 3gluon}(i\pi-\gamma)}
   \left[N\ket{1}-\ket{2}\right]\,.
\lab{B-3g}
\ee
We recognize that this relation between $B_{\rm 3gluon}$ and the
contribution of
three-gluon diagrams, $\Phi_{\rm 3gluon}$, to the cusp anomalous dimension
is exactly the same as that for the QED-like diagrams in \re{B-qed}.

\subsection{Two-loop cross anomalous dimension}

Using the results of calculation of the Feynman diagrams of fig.3--5
we find the two-loop corrections to the line functions $W_1$ and $W_2$
as follows:
$$
W_a^{(2)}=W_a^{\rm QED}+W_a^{\rm self}+W_a^{\rm 3gluon}, \qqquad (a = 1,\ 2)
\,,
$$
where $W_a^{\rm QED}$ was defined in \re{W1(qed)} and \re{W2(qed)},
$W_a^{\rm self}$
in \re{W1(self)} and \re{W2(self)} and $W_a^{\rm 3gluon}$ in \re{W1(3g)} and
\re{W2(3g)}.
Recall, that the two-loop expressions for the Wilson lines
$$
W_a=W_a^{(0)}+W_a^{(1)}+W_a^{(2)}\,, \qqquad (a = 1,\ 2)\,,
$$
should satisfy the RG equation \re{RG}. Here, $W_a^{(0)}$ is the Born terms
defined in \re{Born} and $W_a^{(1)}$ is the one-loop correction
given by \re{W1(1)} and \re{W2(1)}. As a nontrivial test of our calculations we
find after cumbersome calculation that the two-loop Wilson lines $W_1$ and
$W_2$ do obey the RG equation \re{RG}.
Although the coefficients in front of $\log^2(\mu^2/\lambda^2)$ in the
two-loop expressions for the diagrams of fig.3--5 are complicated,
being combined together they give rise to the beta-function term
in the RG equation \re{RG} for $W_a$.

The RG equation \re{RG} implies that the coefficients in front of
$\log(\mu^2/\lambda^2)$ in the two-loop expression for the Wilson line
$W_1^{(2)}$ contribute to the elements $\Gamma_{11}$ and $\Gamma_{12}$
of the matrix of cross anomalous dimensions $\Gamma_{\rm cross}$ and analogous
coefficients in $W_2^{(2)}$ contribute to the elements $\Gamma_{21}$ and
$\Gamma_{22}$. Notice that $W_1^{\rm QED}$ and $W_1^{3gluon}$ do not contain
$\log(\mu^2/\lambda^2)$ terms and the elements $\Gamma_{11}$ and
$\Gamma_{12}$ get contribution only from the self-energy diagrams of fig.4.
Using the expression \re{W1(self)} for $W_1^{\rm self}$ we get
two-loop corrections to the elements of the matrix of cross anomalous
dimension
\baa
\Gamma_{11}^{(2)}(\gamma,g)
&=&-\frac{31}{36}\lr{\alpi}^2 i\pi\coth\gamma
\\
\Gamma_{12}^{(2)}(\gamma,g)&=&\frac{31}{36} \lr{\alpi}^2 i\pi N\coth\gamma\,.
\eaa
These expressions differ from analogous one-loop expressions \re{cross(1)}
only by the factor $\alpi N\frac{31}{36}$. Moreover, there is a simple
relation between $\Gamma_{11}^{(2)}$ and $\Gamma_{12}^{(2)}$ and the two-loop
contribution of the self-energy diagrams to the cusp anomalous dimension
\ba
\Gamma_{11}^{(2)}(\gamma,g)&=&-\frac1{N^2}\lr{\alpi N}^2
                     \lr{\Phi_{\rm self}(\gamma)-\Phi_{\rm self}(i\pi-\gamma)}
\nonumber
\\
\Gamma_{12}^{(2)}(\gamma,g)&=&\frac1{N}\lr{\alpi N}^2
                     \lr{\Phi_{\rm self}(\gamma)-\Phi_{\rm self}(i\pi-\gamma)}
\,.
\lab{cross1}
\ea
Calculating two remaining elements, $\Gamma_{21}^{(2)}$ and
$\Gamma_{22}^{(2)}$, we find that all three different kinds of the diagrams,
$W_2^{\rm QED}$, $W_2^{\rm self}$ and $W_2^{\rm 3gluon}$, contribute to them.
Since the
corresponding $B-$coefficients are related to the cusp anomalous dimension
by the identities \re{B-qed}, \re{B-self} and \re{B-3g}, the final expression
for the elements
of the matrix of cross anomalous dimension can be represented as
\ba
\Gamma_{21}^{(2)}&=&-\frac1{N}\lr{\alpi N}^2
\left[\Phi_{\rm cusp}(\gamma)
 -2\lr{\Phi_{\rm cusp}(\gamma)-\Phi_{\rm cusp}(i\pi-\gamma)}
 +\lr{\Phi_{\rm self}(\gamma)-\Phi_{\rm self}(i\pi-\gamma)}
\right]
\nonumber
\\
\Gamma_{22}^{(2)}&=&\lr{\alpi N}^2
\left[\Phi_{\rm cusp}(\gamma)
 -\frac2{N^2}\lr{\Phi_{\rm cusp}(\gamma)-\Phi_{\rm cusp}(i\pi-\gamma)}
 +\frac1{N^2}\lr{\Phi_{\rm self}(\gamma)-\Phi_{\rm self}(i\pi-\gamma)}
\right]\,,
\nonumber
\\
\lab{cross2}
\ea
where $\Phi_{\rm cusp}=\Phi_{\rm QED}+\Phi_{\rm self}+\Phi_{\rm 3gluon}$
was defined in \re{cusp2}. The terms
containing $\Phi_{\rm self}$ appeared in
the r.h.s. of these relations due to the additional one-half factor in the
expression \re{B-self} for $B_{\rm self}$. The terms containing
$\Phi_{\rm cusp}(\gamma)$
look like two-loop corrections to the cusp anomalous dimension \re{cusp1}
defined in the {\it adjoint\/} representation of the $SU(N)$.

\section{Properties of the cross anomalous dimension}

Having the explicit relations \re{cross1} and \re{cross2} between the
two-loop corrections
to $\Gamma_{\rm cross}(\gamma,g)$ and $\Gamma_{\rm cusp}(\gamma,g)$,
we summarize our two-loop calculations by representing the matrix of cross
anomalous dimensions in the following form:
\ba
\Gamma^{11}_{\rm cross}&=&-\frac1{N^2}\Gamma_1(\gamma,g)
\nonumber
\\
\Gamma^{12}_{\rm cross}&=&\frac1{N}\Gamma_1(\gamma,g)
\lab{main0}
\\
\Gamma^{21}_{\rm cross}&=&-\frac1{N}
\left[\Gamma_{\rm cusp}(\gamma,g)
 +\Gamma_1(\gamma,g)
 +\Gamma_2(\gamma,g)
\right]
\nonumber
\\
\Gamma^{22}_{\rm cross}&=&\Gamma_{\rm cusp}(\gamma,g)
+\frac1{N^2}\left[\Gamma_1(\gamma,g)+\Gamma_2(\gamma,g)\right]\,,
\nonumber
\ea
where the notation was introduced
\ba
\Gamma_1(\gamma,g)&=&
\left[\alpi N+\frac{31}{36}\lr{\alpi N}^2\right] i\pi\coth\gamma
\nonumber
\\
\Gamma_2(\gamma,g)&=&-2\lr{\Gamma_{\rm cusp}(\gamma,g)
                          -\Gamma_{\rm cusp}(i\pi-\gamma,g)}\,.
\lab{main1}
\ea
We note, that the form of this expression was deduced from the properties of
two-loop corrections \re{cross1} and \re{cross2} and it comes as a surprise
that the one-loop
expression \re{cross(1)} for $\Gamma_{\rm cross}(\gamma,g)$ also obeys this
relation. Interesting property of the obtained expression is that it expresses
four originally independent elements of the matrix
$\Gamma_{\rm cross}(\gamma,g)$
in terms of only two functions, $\Gamma_{\rm cusp}$ and $\Gamma_1(\gamma,g)$.
Written in this form, the relation \re{main0} admits natural generalization
to all orders of perturbation theory. Although we do not have a proof
that \re{main0} is valid to all orders of PT, we give below arguments
in favour of \re{main0}.

Let us examine whether the properties of one-loop $\Gamma_{\rm cross}$
found in sect.2 are still valid after we take into account two-loop
corrections in \re{main0}.

\subsection{Imaginary part}

The cusp anomalous dimension $\Gamma_{\rm cusp}(\gamma,g)$ takes real positive
values for real angles $\gamma$ while
$\Gamma_{\rm cusp}(i\pi-\gamma,g)$ has a nontrivial imaginary part.
It turns out that in the expression for the cross anomalous dimension
\re{main0}, $\Gamma_1(\gamma,g)$ and $\Gamma_2(\gamma,g)$
have pure imaginary values and change sign under reflection
$\gamma\to i\pi-\gamma$:
$$
\lr{\Gamma_a(\gamma,g)}^\dagger=-\Gamma_a(\gamma,g),
\qquad
\Gamma_a(i\pi-\gamma,g)=-\Gamma_a(\gamma,g),
\qquad (a=1,\ 2)\,.
$$
This property is obvious for the function
$\Gamma_1(\gamma,g)$ defined in \re{main1}. To prove it for
the function $\Gamma_2(\gamma,g)$, we apply the following property of
the cusp anomalous dimension
\baa
2\Gamma_{\rm cusp}(\gamma,g)&=&\Gamma_{\rm cusp}(i\pi-\gamma,g)
                        +\Gamma_{\rm cusp}(-i\pi-\gamma,g)
\\                      &=&\Gamma_{\rm cusp}(i\pi-\gamma,g)
                        +\left[\Gamma_{\rm cusp}(i\pi-\gamma,g)\right]^\dagger
\eaa
which can be easily verified using \re{cusp1}.%
\footnote{In fact, the integrals $I_1$, $I_2$ and $I_3$ defined in \re{I1-I3}
          satisfy analogous relation.}
As a result, the function $\Gamma_2(\gamma,g)$ can be expressed as
\baa
\Gamma_2(\gamma,g)=2i\ \Im \Gamma_{\rm cusp}(i\pi-\gamma,g)
&=&\Gamma_{\rm cusp}(i\pi-\gamma,g)-\Gamma_{\rm cusp}(-i\pi-\gamma,g)
\\
&=&\Gamma_{\rm cusp}(i\pi-\gamma,g)-\Gamma_{\rm cusp}(i\pi+\gamma,g)
\eaa
where in the last identity we used the fact that $\Gamma_{cusp}(\gamma,g)$
is an even function of the angle $\gamma$.
Let us calculate the determinant of the matrix \re{main0}:
$$
\det\Gamma_{\rm cross}(\gamma,g)=\frac{N^2-1}{N^4}
\Gamma_1(\gamma,g) (\Gamma_1(\gamma,g)+\Gamma_2(\gamma,g))\,.
$$
We notice that $\Gamma_{\rm cusp}$ does not contribute to this expression
due to  special form of the matrix \re{main0}. Although the matrix
$\Gamma_{\rm cross}$ is complex the determinant turns out to be real even
function of the angle $\gamma$, since both $\Gamma_1$ and $\Gamma_2$ are
imaginary odd functions of $\gamma$.

\subsection{Large $\gamma$ asymptotics}

Let us consider the asymptotic behavior of the cross anomalous dimension
for large angles $\gamma$ between velocities $v_1$ and $v_2$. We recall
that $\gamma$ is defined as an angle in Minkowski space-time and it
may have arbitrary values. The large $\gamma$ limit \re{gl}
corresponds to the high energy scattering of quarks with
velocities $v_1$ and $v_2$.

To find the asymptotics of the expression \re{main0} for $\gamma\gg 1$ we
need to know the analogous behavior of the functions
$\Gamma_{\rm cusp}(\gamma,g)$ and $\Gamma_{\rm cusp}(i\pi-\gamma,g)$.
To this end we study the most general situation and consider
$\Gamma_{\rm cusp}(\chi,g)$ with $\chi$ being the complex quantity
with large (negative or positive) real part: $(\Re \chi)^2\gg 1$.
After some algebra we get from the two-loop expression \re{cusp1} for the cusp
anomalous dimension that $\Gamma_{\rm cusp}(\chi,g)$ is a
linear function of $\chi$ in this limit,
\be
\Gamma_{\rm cusp}(\chi,g) = \chi\ \sign(\Re\chi)\ \Gamma_{\rm cusp}(g)
+ \CO((\Re \chi)^0)\,,
\lab{as}
\ee
where the two-loop expression for the coefficient $\Gamma_{cusp}(g)$ is
given by%
\footnote{We recall, that $\Gamma_{\rm cusp}$ is defined in
          the adjoint representation of the $SU(N)$ group.}
$$
\Gamma_{\rm cusp}(g)=
\alpi N + \lr{\alpi N}^2\lr{\frac{67}{36}-\frac{\pi^2}{12}}
\,.
$$
Moreover, as was shown in \ci{cusp}, this asymptotic behavior is valid
to all orders of PT. The higher order corrections to
$\Gamma_{\rm cusp}(\gamma,g)$ only modify the coefficient
$\Gamma_{\rm cusp}(g)$.
Using the relations \re{as} and \re{main1} we get the asymptotics of the
function $\Gamma_2$ for large positive angles $\gamma$ as
$$
\Gamma_2(\gamma,g)=-2 i\pi \Gamma_{\rm cusp}(g) + \CO(\gamma^{-1})
$$
and the analogous expression for the function $\Gamma_1$ is
$$
\Gamma_1(\gamma,g)= i\pi \Gamma_1(g) + \CO(\gamma^{-1}),\qquad
\Gamma_1(g)=\alpi N +\frac{31}{36} \lr{\alpi N}^2\,.
$$
Thus, the large $\gamma$ asymptotic behavior of the matrix of cross
anomalous dimensions \re{main0} is
\ba
\Gamma^{11}_{\rm cross}&=&-\frac{i\pi}{N^2}\Gamma_1(g)
\nonumber
\\
\Gamma^{12}_{\rm cross}&=&\frac{i\pi}{N}\Gamma_1(g)
\lab{cross-as}
\\
\Gamma^{21}_{\rm cross}&=&-\gamma\frac1{N}\Gamma_{\rm cusp}(g)
 +\frac{i\pi}{N}\lr{2\Gamma_{\rm cusp}(g)-\Gamma_1(g)}
\nonumber
\\
\Gamma^{22}_{\rm cross}&=&\gamma\Gamma_{\rm cusp}(g)
-\frac{i\pi}{N^2}\lr{2\Gamma_{\rm cusp}(g)-\Gamma_1(g)}\,,
\nonumber
\ea
where we omitted nonleading in $\gamma$ terms in real and imaginary parts
of the matrix elements.

Let us calculate the eigenvalues $\Gamma_\pm$ of the matrix
$\Gamma_{\rm cross}(\gamma,g)$. From \re{cross-as} we find their sum as
$$
\Gamma_+ + \Gamma_- = \tr \Gamma_{\rm cross}(\gamma,g)=
\lr{\gamma-\frac{2i\pi}{N^2}}\Gamma_{\rm cross}(g)
$$
and we may expect that both eigenvalues are of order $\gamma$. However,
calculating the determinant of \re{cross-as} we find
the product of eigenvalues behaves as $\gamma^0$,
$$
\Gamma_+ \Gamma_- = \det \Gamma_{\rm cross}(\gamma,g) =
\pi^2\frac{N^2-1}{N^4} \Gamma_1(g)(2\Gamma_{\rm cusp}(g)-\Gamma_1(g))
\,.
$$
This means
that one of the eigenvalues is much larges than the second
one: $\Gamma_+=\CO(\gamma)$ and $\Gamma_-=\CO(\gamma^{-1})$.
The explicit expressions for the eigenvalues are:
\ba
\Gamma_+ &=&\lr{\gamma-\frac{2i\pi}{N^2}}\Gamma_{\rm cusp}(g)
\lab{values}
\\
\Gamma_- &=&\lr{\frac1{\gamma}+\frac{2i\pi}{N^2}\frac1{\gamma^2}}
\pi^2\frac{N^2-1}{N^4}\frac{\Gamma_1(g)(2\Gamma_{\rm cusp}(g)-\Gamma_1(g))}
                            {\Gamma_{\rm cusp}(g)}\,,
\nonumber
\ea
where we omitted nonleading in $\gamma$ terms in real and imaginary parts
of these expressions.

To get some insight into the general structure of the matrix
$\Gamma_{\rm cross}$, let us study the large $N$ limit defined as follows:
$$
\alpha_s N = \mbox{fixed} , \qquad N \to \infty\,.
$$
We find from \re{cusp1}, \re{main1} and \re{main0}
that the anomalous dimensions $\Gamma_{\rm cusp}(\gamma,g)$,
$\Gamma_1(\gamma,g)$ and $\Gamma_2(\gamma,g)$ survive in this limit.
At the same time, all the elements of the matrix $\Gamma_{cross}$ vanish
except of the element $\Gamma_{22}$ which becomes equal to the cusp anomalous
dimension. As we show in Appendix C, this observation is valid to all loop
order in PT and the elements of the matrix of cross anomalous dimension
have the following large $N$ behavior
\be
\Gamma^{11}_{\rm cross} = \CO(N^{-2}), \qquad
\Gamma^{12}_{\rm cross} = \CO(N^{-1}), \qquad
\Gamma^{21}_{\rm cross} =\CO(N^{-1}), \qquad
\Gamma^{22}_{\rm cross} =\Gamma_{\rm cusp}(\gamma,g)+\CO(N^{-2})\,,
\lab{N0}
\ee
where $\Gamma_{\rm cusp}(\gamma,g)$ is defined in the adjoint representation
of the $SU(N)$. We notice that the expression \re{main0} for the cross
anomalous dimension to higher orders of PT is consistent with large $N$
behavior \re{N0}.

Summarizing our two-loop calculations we conclude that the higher order
corrections to the matrix of anomalous dimensions are organized in such a
way, that they preserve the asymptotic behavior of the eigenvalues
\re{values}. This allows us to apply the results of sect.3.2 and find
the higher energy behavior of the scattering amplitudes in QCD.

\section{Conclusions}

In this paper we considered the elastic quark-quark scattering at high
energy $s$ and fixed transferred momentum $t$. We have shown that
all effects of interaction of incoming quarks with soft gluons are factorized
into Wilson lines which appear as eikonal phases of the scattering quarks.
As a consequence, the quark-quark scattering amplitude is represented as
an expectation value of a Fourier transformed Wilson line
evaluated along the integration path which consists of two semiclasical
quark trajectories separated by the impact parameter in the transverse
direction. We evaluated the scattering amplitude using the fact
that for zero impact parameter the Wilson line has additional cross
singularities. We have shown that the asymptotic behavior
of the scattering amplitude is governed by the matrix of cross anomalous
dimension $\Gamma_{\rm cross}(\gamma,g)$ which depends on the energy
of incoming quarks through the angle $\gamma$ between quark velocities.
We performed two-loop calculation of $\Gamma_{\rm cross}(\gamma,g)$ and
found remarkable properties of this matrix. It turned out that four
originally independent elements of $\Gamma_{\rm cross}(\gamma,g)$ can be
expressed in terms of only two functions, the cusp anomalous dimension
$\Gamma_{\rm cusp}(\gamma,g)$ and $\Gamma_1(\gamma,g)$ which comes from
self-energy diagrams. Written in this form, the elements of
$\Gamma_{\rm cross}(\gamma,g)$ admit natural generalization to higher
orders of PT. The form of the corresponding expressions was confirmed
by the large $N$ analysis of the Wilson lines. Higher order corrections
are organized in such a way that they preserve the asymptotic behavior
of the eigenvalues of the matrix of cross anomalous dimension.
This allowed us to apply the one-loop result for
$\Gamma_{\rm cross}(\gamma,g)$ to find the high energy asymptotic of
the quark-quark scattering amplitude. This amplitude can be decomposed
into singlet and octet invariant amplitudes corresponding to the exchange
in the $t-$channel with quantum numbers of vacuum and gluon, respectively.
To any order of PT theory the octet amplitude gets large double
logarithmic corrections while corrections to the singlet amplitude
are nonleading. However, after resummation to all orders of PT
these large corrections are summed into the exponentially small
Sudakov like exponent with the slope depending on the eigenvalues of the
matrix of cross anomalous dimensions. As a result, with leading and
nonleading logarithmic corrections taken into account, the singlet amplitude
dominates in the high energy asymptotic behavior of the quark-quark
scattering amplitude.

\bigskip\par\noindent{\Large {\bf Acknowledgements}}\par\bigskip\par

\noindent
We are grateful for helpful conversations with J.~De Boer, G.~Marchesini,
G.~Sterman and A.~Radyushkin.
This work was supported in part by the National Science Foundation
under grant PHY9309888.


\appendix{A}{Basic diagrams}
Let us consider the diagram of fig.3(1). Using the parameterization of the
integration path \re{para} we find the following expression for the Feynman
integral:
$$
F_1^{\rm QED}
=(ig)^4(v_1v_2)^2
\int_0^\infty d\alpha_1 \int_0^{\alpha_1} d\alpha_2 \int_0^\infty d\alpha_3
\int_{-\infty}^0 d\beta\
D(v_1\alpha_1-v_2\beta)D(v_1\alpha_2-v_2\alpha_3)\,,
$$
where the $D-$function was defined in \re{prop}.
It is convenient to introduce the following
integrals:
\ba
I_1(\gamma)&=&\int_0^\gamma d\psi\ \psi\coth\psi
\nonumber
\\
I_2(\gamma)&=&\int_0^\gamma d\psi\ \psi(\gamma-\psi)\coth\psi
\lab{I1-I3}
\\
I_3(\gamma)&=&\int_0^\gamma d\psi\ \frac{\psi\coth\psi-1}
                                    {\sinh^2\gamma-\sinh^2\psi}
\log\frac{\sinh\gamma}{\sinh\psi}\,.
\nonumber
\ea
Thus defined functions obey the following property
\be
I_2(i\pi-\gamma)+\gamma I_1(i\pi-\gamma)
=I_2(\gamma)+(i\pi-\gamma) I_1(\gamma)\,.
\lab{prope}
\ee
Then the final expression for the Feynman integral looks like
\baa
F_1^{\rm QED}(\gamma)&=&\frac14\lr{\alpi}^2
            \lr{\frac{4\pi\mu^2}{\lambda^2}}^{2\eps}
\Gamma(2\eps)\coth^2\gamma
\\
&\times&
\left(
\frac{\Gamma(1+\eps)\Gamma(\eps)}{\Gamma(1+2\eps)}
\gamma(i\pi-\gamma)
+2\gamma I_1(i\pi-\gamma)   
-2(i\pi-\gamma) I_1(\gamma) 
\right)\,.
\eaa
The cross singularities appear in $F_1$ as a double pole in $\eps$.
However, the
diagram contains a divergent subgraph and in order to renormalize
$F_1$ we have to subtract the cross divergence of this subgraph from $F_1$.
The corresponding counter term is proportional in the $\MS-$scheme to the
one-loop expression \re{F1}
$$
\left[F_1^{\rm QED}(\gamma)\right]_{c.-t.}=-\lr{\alpi}^2
\lr{\frac{4\pi\mu^2}{\lambda^2}}^\eps \frac{\Gamma(1+\eps)}{4\eps^2}
\gamma(i\pi-\gamma)\coth^2\gamma\,.
$$
To renormalize the contribution of the diagram we add
$\left[F_1^{\rm QED}\right]_{c.-t.}$
to $F_1^{\rm QED}$ and subtract poles from the sum in the $\MS-$scheme to find
the following renormalized expression
$$
F_1^{\rm QED}(\gamma)=\lr{\alpi}^2\coth^2\gamma
\left\{\frac18\gamma(i\pi-\gamma)\log^2\frac{\mu^2}{\lambda^2}
+\frac12\left[
 \gamma I_1(i\pi-\gamma)-(i\pi-\gamma) I_1(\gamma)
    \right]\log\frac{\mu^2}{\lambda^2}
\right\}\,.
$$
The calculation of the diagram of fig.3(2) is analogous to that of
$F_1^{\rm QED}$
and the final renormalized expression for the diagram is
$$
F_2^{\rm QED}(\gamma)=\lr{\alpi}^2\coth\gamma
\left\{\frac18\gamma\log^2\frac{\mu^2}{\lambda^2}
+\frac12\left[I_1(\gamma)-\gamma\right]\log\frac{\mu^2}{\lambda^2}
\right\}
$$
where the corresponding counter term was taking into account.

\subsection{Cusp anomalous dimension}

The diagrams of fig.3(3)-(5), fig.4(1) and fig.5(1) have been already
calculated in \ci{cusp}.  Using the results of the paper \ci{cusp} we
get the following renormalized expressions for the diagram of fig.3(3)
$$
F_3^{\rm QED}=\lr{\alpi}^2I_2(\gamma)\ \coth^2\gamma\
\log\frac{\mu^2}{\lambda^2}
\,,
$$
where the integral $I_2$ was defined in \re{I1-I3}.
For the diagram of fig.3(4) we have
$$
F_4^{\rm QED}=\lr{\alpi}^2\coth\gamma
\left\{\frac18\gamma\log^2\frac{\mu^2}{\lambda^2}
+\frac12\left[\gamma-I_1(\gamma)\right]\log\frac{\mu^2}{\lambda^2}
\right\}\,,
$$
for the diagram of fig.3(5) we get
$$
F_5^{\rm QED}=\lr{\alpi}^2\coth^2\gamma
\left[\frac18\gamma^2\log^2\frac{\mu^2}{\lambda^2}
-I_2(\gamma)\log\frac{\mu^2}{\lambda^2}
\right]\,,
$$
for the diagram of fig.4(1) containing self-energy correction to the
gluon propagator
$$
F_1^{\rm self}=-\lr{\alpi}^2\lr{\frac5{48}\log^2\frac{\mu^2}{\lambda^2}
                  +\frac{31}{72}\log\frac{\mu^2}{\lambda^2}}
            \gamma\coth\gamma
\,,
$$
for the diagram of fig.5(1)
$$
F_1^{\rm 3gluon}=\lr{\alpi}^2
\left[
\frac{\pi^2}{96}(\gamma\coth\gamma-1)+\frac18 I_3(\gamma)\sinh(2\gamma)
\right]\log\frac{\mu^2}{\lambda^2}
\,.
$$
As was shown in \ci{cusp}, all these diagrams contribute to the two-loop cusp
anomalous dimension.

\subsection{Ward identities}

The Feynman diagrams of figs.3(7) and 3(4) can be interpreted as propagator and
vertex corrections to the one-loop diagram of fig.2(1). As was shown in
\ci{cusp},
there are the Ward identities for the Wilson lines, analogous to the Ward
identities between Green functions in QCD, which relate these two diagrams
as follows:
$$
F_7^{\rm QED}=-F_4^{\rm QED}\,.
$$
The same identity is valid for the diagrams of figs.4(8) and 4(7):
$$
F_8^{\rm QED}=-F_7^{\rm QED}=F_4^{\rm QED}\,,
$$
and for the diagrams of figs.5(2) and 5(1):
$$
F_2^{\rm self}=-\frac12 F_1^{\rm self}(0)\,,
$$
where the factor one-half takes into account that the self-energy correction
is a ``square root'' of the propagator and $F_1^{\rm self}(0)$ means the
value of the Feynman integral $F_1^{\rm self}$ evaluated at $\gamma=0$.

\subsection{Reflection symmetry: $\gamma\to i\pi-\gamma$}

The integration paths in figs.1(a) and 1(b) are defined by two vectors $v_1$
and
$v_2$ in Minkowski space-time. Let us consider the properties of the diagrams
under the reflection $v_2\to -v_2$ or in terms of the angle
between vectors $\gamma\to i\pi-\gamma$. Under this transformation the
integration path transforms into itself but the orientation of the $v_2-$line
becomes opposite to the original one. After inverting of the direction of
integration along $v_2-$line as $\int_a^b dx A(x)=-\int_b^a dx A(x)$
we get formally another diagram multiplied by the factor $(-)^\#$
with $\#$ being the number of gluons attached to the $v_2-$line. Thus,
taking one of the diagrams of fig.3, we change the angle in the
corresponding Feynman integral as $\gamma\to i\pi-\gamma$ and multiply
the result by appropriate $(-)^\#$ to get the Feynman integral corresponding
to another diagram. As a result, we find the following useful
relations between the Feynman integrals of QED like diagrams
$$
    F_{9}^{\rm QED} = F_3^{\rm QED}(i\pi-\gamma)
,\qquad
    F_{10}^{\rm QED} = -F_4^{\rm QED}(i\pi-\gamma)
,\qquad
    F_{11}^{\rm QED} =  F_5^{\rm QED}(i\pi-\gamma)
$$
$$
    F_{12}^{\rm QED} = -F_7^{\rm QED}(i\pi-\gamma)= F_4^{\rm QED}(i\pi-\gamma)
,\qquad
    F_{13}^{\rm QED} =  F_1^{\rm QED}(i\pi-\gamma)
$$
$$
    F_{14}^{\rm QED} = -F_8^{\rm QED}(i\pi-\gamma)=-F_4^{\rm QED}(i\pi-\gamma)
,\qquad
    F_{15}^{\rm QED} = -F_2^{\rm QED}(i\pi-\gamma)\,,
$$
and between self-energy and three gluon diagrams
$$
    F_3^{\rm self}   = -F_1^{\rm self}(i\pi-\gamma)
,\qquad
    F_2^{\rm 3gluon} = -F_1^{\rm 3gluon}(i\pi-\gamma)\,.
$$
Notice that we don't have the same relations between the
color factors of the diagrams, because after inverting the direction of the
integration along $v_2-$line the color indices of the gauge fields
$A_\mu(x)=A_\mu^a(x)t^a$ become anti-path-ordered along the $v_2-$line.

\subsection{$\gamma\to 0$ limit}

Let us consider the diagram of fig.3(16) in which all gluons are attached to
only one of the lines. A simple observation is that the Feynman integral of
this diagram can be obtained from the Feynman integral of the diagram of
fig.3(3) in the limit $v_1\to v_2$ or in terms of the angle $\gamma\to 0$.
The same property leads to the following relations between QED like diagrams:
$$
   F_{16}^{\rm QED} =  F_3^{\rm QED}(0), \qquad
   F_{17}^{\rm QED} =  F_4^{\rm QED}(0), \qquad
   F_{18}^{\rm QED} =  F_5^{\rm QED}(0), \qquad
   F_{19}^{\rm QED} =  F_7^{\rm QED}(0)
$$
and between self-energy diagrams
$$
F_4^{\rm self}   =  F_1^{\rm self}(0)\,,
$$
where in the r.h.s.\ the corresponding Feynman integrals are evaluated at
$\gamma=0$.

\subsection{Factorization}

We notice, that in the Feynman diagram of fig.3(20) two gluon propagator are
independently integrated along the path. As a consequence, the corresponding
Feynman integral can be factorized into the product of two independent
one-loop integrals which have been calculated in sect.2.
The same factorization property leads to the following relations:
$$
F_{20}^{\rm QED}=(F_2^{\rm 1-loop})^2,\qquad
F_{21}^{\rm QED}=(F_1^{\rm 1-loop})^2,\qquad
F_{22}^{\rm QED}=F_{23}^{\rm QED}=F_1^{\rm 1-loop}F_4^{\rm 1-loop}
$$
$$
F_{24}^{\rm QED}=F_{25}^{\rm QED}=F_2^{\rm 1-loop}F_4^{\rm 1-loop},\qquad
F_{26}^{\rm QED}=F_{27}^{\rm QED}=F_1^{\rm 1-loop}F_3^{\rm 1-loop},\qquad
F_{28}^{\rm QED}=(F_3^{\rm 1-loop})^2\,.
$$
Using the expressions for the basic Feynman integrals and applying all these
relations we can find the Feynman integrals for
all the diagrams of figs.3--5 except of that for the diagram of fig.5(3).
The calculation of the diagram of fig.5(3) is based on the following
property. It turns out that the sum of the Feynman
integrals corresponding to the diagrams of figs.5(1)-(3) vanishes,
that is
\be
F_3^{\rm 3gluon}=-F_1^{\rm 3gluon}-F_2^{\rm 3gluon}
            =-F_1^{\rm 3gluon}(\gamma)+F_1^{\rm 3gluon}(i\pi-\gamma)\,.
\lab{3g}
\ee
To show this we notice that three diagrams of fig.5 correspond to all
possible positions of two gluons on one of the lines provided that
the gluons are ordered with respect to each other. Moreover,
applying the reflection symmetry we may extend the integration over position of
third gluon from $[0,\infty)$ to $(-\infty,\infty)$ and take one-half
of the result. The calculation of the Feynman integrals is more simple
in the momentum space. Let us denote the momenta of gluons as $k_a$,
$a=1,2,3$ with $k_1+k_2+k_3=0$ and $k_3$ being the momentum of a single
gluon attached to the $v_2$ line. Then, taking into account the
expression for three-gluon vertex, we get the following momentum integral
for the sum of the diagrams of fig.5(1)--(3),
$$
\propto (v_1^2 v_2^\mu-(v_1\cdot v_2) v_1^\mu)
        \int \frac{d^D k_3}{k_3^2}\delta(k_3\cdot v_1)\delta(k_3\cdot v_2)
        \int \frac{d^D k_1}{k_1^2}\frac{k_1^\mu}{(k_1+k_3)^2 (k_1\cdot v_1)}
$$
where two delta functions appear after integration over positions of gluons
on $v_1$ and $v_2$ lines from $-\infty$ to $\infty$. The momentum integral
is proportional to the only vector involved, $v_1^\mu$, while the prefactor
is orthogonal to $v_1$. The product is zero which means that the sum of the
Feynman integrals corresponding to the diagrams of fig.5(1)-(3) vanishes.

\appendix{B}{Color factors and combinatorial weights}
The relation
\re{dec} implies that the color factors for the diagrams of fig.3--5 have the
following general form
\be
C = C_1 \delta_{ii'} \delta_{jj'} + C_2 \delta_{ij'} \delta_{ji'}
  = C_1 \ket{1} + C_2 \ket{2}\,,
\lab{gen}
\ee
where the coefficients $C_1$ and $C_2$ are different for the same diagram
contributing to $W_1$ and $W_2$. Recall, that the relations \re{dec} and
\re{gen} are unique properties of the fundamental representation of the $SU(N)$
group.

\subsection{QED like diagrams}
The general form of the color
factors is given by \re{gen} in the fundamental representation of the $SU(N)$.
Consider, as an example, calculation of the color factor of the diagram
of fig.3(1). This diagram contributes to the both line functions,
$W_1$ and $W_2$, with the color factors $C_1^1$ and $C_1^2$,
respectively, given by
\baa
C_1^1&=&(t^a t^b)_{i'i} (t^b t^a)_{j'j}
   =\frac{1}{4N^2}\ket{1}+\frac{N^2-2}{4N} \ket{2}
\\
C_1^2&=&(t^a t^b t^a)_{i'j} t^a_{j'i}
   =-\frac1{4N}\ket{1}+\frac1{4N^2}\ket{2}\,,
\eaa
where we substituted the identity \re{dec} and
the states $\ket{1}$ and $\ket{2}$ were defined in
\re{Born}.
The calculation of the color factors of the remaining QED like diagrams
of fig.3 is analogous.
The calculation of the color factors of the QED like diagrams of fig.3 leads
to the following expressions
$$
C_2^1=C_4^1=C_{10}^1=C_{15}^1=\frac1{4N^2}\ket{1}-\frac1{4N}\ket{2}
$$
$$
C_3^1=C_{11}^1=C_{13}^1=C_{20}^1=\frac{N^2+1}{4N^2}\ket{1}-\frac1{2N}\ket{2}
$$
$$
C_7^1=C_8^1=C_{12}^1=C_{14}^1=C_{22}^1=C_{23}^1=C_{24}^1=C_{25}^1=
-\frac{N^2-1}{4N^2}\ket{1}+\frac{N^2-1}{4N}\ket{2}
$$
$$
C_{16}^1=C_{17}^1=C_{18}^1=C_{19}^1=C_{26}^1=C_{27}^1=C_{28}^1=
-\frac{N^2-1}{4N^2}\ket{1}
$$
$$
C_6^1=N\ket{1}, \qquad C_5^1=C_{21}^1=
\frac1{4N^2}\ket{1}+\frac{N^2-2}{4N}\ket{2}\,,
$$
where the lower index numerates the diagrams of fig.3 and the upper index
is related to the Wilson line $W_1$. The color factors of the diagrams
contributing to the Wilson line $W_2$ are
$$
C_2^2=C_{12}^2=C_{13}^2=C_{19}^2=C_{24}^2=C_{25}^2=C_{26}^2=C_{27}^2=
\frac{N^2-1}{4N}\ket{1}-\frac{N^2-1}{4N^2}\ket{2}
$$
$$
C_3^2=C_4^2=C_5^2=C_6^2=C_7^2=C_{21}^2=C_{22}^2=C_{23}^2
=-\frac{N^2-1}{4N^2}\ket{2}
$$
$$
C_8^2=C_{10}^2=C_{17}^2=-\frac1{4N}\ket{1}+\frac1{4N^2}\ket{2}
$$
$$
C_9=C_{15}=C_{18}=C_{28}=\frac{N^2-2}{4N}\ket{1}+\frac1{4N^2}\ket{2}
$$
$$
C_{11}^2=C_{14}^2=C_{16}^2=C_{20}^2
        =-\frac1{2N}\ket{1}+\frac{N^2+1}{4N^2}\ket{2}\,.
$$
The combinatorial weight of the QED like diagrams are given by
$$
w_3=w_5=w_9=w_{11}=w_{16}=w_{18}=2, \qquad
w_{20}=w_{21}=w_{28}=1
$$
and the combinatorial weight of the remaining diagrams of fig.3 is equal to 4.

\subsection{Self-energy diagrams}

The color factors of the self-energy diagrams of fig.4 are defined as follows:
$$
C_1^1=C_3^1=-\frac12\ket{1}+\frac{N}2\ket{2}, \qquad
C_2^1=C_4^1=\frac{N^2-1}{2}\ket{1}
$$
$$
C_1^2=C_2^2=\frac{N^2-1}{2}\ket{2}, \qquad
C_3^2=C_4^2=\frac{N}2\ket{1}-\frac12\ket{2}
$$
and the combinatorial weights of the diagrams are
$$
w_1=w_3=w_4=2,\qquad w_2=4\,.
$$

\subsection{Three-gluon diagrams}
The diagrams of fig.5 contribute to the line function $W_1$ with the
color factors
$$
C_1^1=C_2^1=C_3^1=-2if^{abc}(t^a t^b)_{i'i} t^c_{j'j}
           =-\frac12\ket{1}+\frac{N}2\ket{2}\,,
$$
where $f^{abc}$ are structure constants of the $SU(N)$ and the combinatorial
weights are
$$
w_1=w_2=w_3=4 .
$$
For the line function $W_2$  combinatorial weights are the same but the color
factors of the diagrams of fig.5 are different
$$
C_1^2=\frac{N^2-1}2\ket{2}, \qquad
C_2^2=-C_3^2=\frac{N}2\ket{1}-\frac12\ket{2}\,.
$$

\appendix{C}{Large $N$ limit of the cross anomalous dimension}
To find relations \re{N0}
we consider the RG equation \re{RG} for the
Wilson lines $W_a$ and multiply the both sides of the equation by
$\delta_{i'i}\delta_{j'j}$. After contacting of the color indices we get
from $W_1$ and $W_2$ two Wilson loops which are
closed at infinity and have the following large
$N$ normalization
\be
(W_1)^{i'j'}_{ij}\delta_{i'i}\delta_{j'j}= N^2  w_1, \qquad
(W_2)^{i'j'}_{ij}\delta_{i'i}\delta_{j'j}= N    w_2, \qquad
\lab{w}
\ee
with $w_a=\CO(N^0)$ and $a=1,\ 2$. We notice, that such defined $w_1$
is given by the vacuum averaged product of two ``elementary'' Wilson loops
calculated along $v_1$ and $v_2$ lines. The simplification of $w_1$ appears
in the large $N$ limit after we will take into account the following
factorization property (vacuum dominance) \ci{Mig}
$$
\langle 0 | \CO_1 \CO_2 |0  \rangle  =
\langle 0 | \CO_1 |0 \rangle \langle 0 | \CO_2 |0 \rangle
\times \left(1+\CO(N^{-2})\right)
$$
valid for two arbitrary gauge invariant operators $\CO_1$ and $\CO_2$. In our
case, $\CO_1$ is given by $P\exp(ig\int_{-\infty}^\infty ds v_1 A(v_1s)$
and the expression for $\CO_2$ is analogous with replacement $v_1$ by $v_2$.
As a result, the Wilson loop $w_1$ has a trivial value
\be
w_1 = 1 + \CO(N^{-2}).
\lab{tri}
\ee
since $\langle0|\CO_1|0\rangle=\langle0|\CO_2|0\rangle=1$.
Notice, that the first nonleading correction to $w_1$ behaves as
$\CO(N^{-2})$ and not like $\CO(N^{-1})$.

After substitution of \re{w} and \re{tri} into the RG equation \re{RG} we
find the following identity in the large $N$ limit
$$
\Gamma^{11}_{\rm cross}+\frac1{N}\Gamma^{12}_{\rm cross}w_2 = \CO(N^{-2}) \,.
$$
The requirement that the both terms in the l.h.s.\
of this relation have the same large $N$ behavior leads together
with $w_2=\CO(N^0)$ to the conditions
\be
\Gamma^{11}_{\rm cross} = \CO(N^{-2}), \qquad
\Gamma^{12}_{\rm cross} = \CO(N^{-1})\,,
\lab{N1}
\ee
which are in accordance with \re{main0}.
Let us consider the projection of the Wilson lines $W_1$ and $W_2$ onto the
color structure $\delta_{i'j}\delta_{j'i}$
\be
(W_1)^{i'j'}_{ij}\delta_{i'j}\delta_{j'i}= N  w_1', \qquad
(W_2)^{i'j'}_{ij}\delta_{i'i}\delta_{j'j}= N^2  w_2', \qquad
\lab{w'}
\ee
and repeat the large $N$ analysis for the Wilson loops $w_1'$ and $w_2'$.
After substitution of \re{w'} into the RG equation \re{RG} we get
\be
\CD w_2' = \frac1{N} \Gamma^{21}_{\rm cross}w_1'+\Gamma^{22}_{\rm cross}w_2'\,,
\lab{nn}
\ee
where $\CD$ denotes the RG operator.
The Wilson loop $w_1'$ satisfies analogous relation.
As in the previous case, the Wilson loops $w_2'$ is factorized in
the large $N$ limit into the product of two vacuum averaged Wilson loops
evaluated along the paths having a cusp with the same angle $\gamma$
introduced before. As a result, in the large $N$ limit $w_2'$ has cusp
singularities which are renormalized multiplicatively. This means, that,
up to $\CO(N^{-2})$ corrections, $w_2'$ should obey the homogeneous RG
equation with the anomalous dimension
$\frac{2C_F}{C_A}\Gamma_{\rm cusp}(\gamma,g)$
equal to the cusp anomalous dimension
defined in the {\it fundamental\/} representation
of the $SU(N)$ times the number of cusps $(=2)$
with $C_F=(N^2-1)/(2N)$ and $C_A=N$. Since the Wilson loops
$w_1'$ and $w_2'$ behave as $\CO(N^0)$, this condition together with
\re{nn} implies that
\be
\Gamma^{21}_{\rm cross}=\CO(N^{-1}), \qquad
\Gamma^{22}_{\rm cross}=\Gamma_{\rm cusp}(\gamma,g)+\CO(N^{-2})
\lab{N2}
\ee
with $\Gamma_{\rm cusp}(\gamma,g)$ defined in the adjoint representation
of the $SU(N)$. Thus, in the large $N$ limit the matrix
$\Gamma_{\rm cross}(\gamma,g)$ is given by \re{N1} and \re{N2}
to all orders of PT.

\newpage

\bb{99}

\bi{Regge}
      P.D.B. Collins, {\it ``An introduction to Regge theory and high energy
      physics''\/}, Cambridge Univ. Press (1977) 445.
\bi{BFKL}
      E.A. Kuraev,  L.N. Lipatov and V.S. Fadin, Sov.Phys.JETP
      44 (1976) 443-451; 45 (1977) 199;
\\    Ya.Ya. Balitskii and L.N. Lipatov, Sov.J.Nucl.Phys. 28 (1978) 822.
\bi{GLLA}
      H. Cheng and T.T. Wu, {\it ``Expanding Protons: Scattering at
      High Energies''\/}, (MIT Press, Cambridge, Massachusetts, 1987).
\bi{Bar}
      J. Bartels,Nucl. Phys. B175 (1980) 365.
\bi{two-dim}
      L.N. Lipatov, Nucl. Phys. B365 (1991) 614.
\bi{qua}
      L.N. Lipatov, Nucl. Phys. B309 (1988) 379;
      in {\it ``Perturbative QCD''\/}, ed. A.H. Mueller
      (World Scientific, Singapore, 1989).
\bi{Lip}
      L.N. Lipatov, Padova preprint DFPD-93-TH-70; JETP Lett. 59 (1994) 596.
\bi{FK}
      L.D. Faddeev and G.P. Korchemsky, Stony Brook preprint, ITP-SB-94-14,
      hep-th@xxx/9404173
\bi{Ver}
      H. Verlinde and E. Verlinde, Princeton Univ. preprint
      PUPT-1319, hep-th@xxx/9302104.
\bi{evol}
      M.G. Sotiropoulos and G. Sterman, Nucl. Phys. B419 (1994) 59.
\bi{K}
      G.P. Korchemsky, Phys. Lett. B325 (1994) 459.
\bi{Mig}
      A.A. Migdal, Phys. Rep. 102 ( 1983) 199.
\bi{pQCD}
      G.P. Korchemsky and A.V. Radyushkin,
      Sov. J. Nucl. Phys. 44 (1986) 145; 45 (1987) 127; 910;
      Phys. Lett. B171 (1986) 459; B279 (1992) 359;
\\    G.P. Korchemsky, Phys. Lett. 217B (1989) 330; 220B (1989) 629;
      Mod. Phys. Lett. A4 (1989) 1257.
\bi{QED}
      H. Cheng and T.T. Wu, Phys. Rev. Lett. 22 (1969) 666;
\\    H. Abarbanel and C. Itzykson,  Phys. Rev. Lett. 23 (1969) 53.
\bi{Br}
      R.A. Brandt, F. Neri and M.-A. Sato, Phys. Rev. D24 (1981) 879;
\\    R.A. Brandt, A. Gocksch, M.-A. Sato and F. Neri,
      Phys. Rev. D26 (1982) 3611.
\bi{Pol}
      A.M. Polyakov, Nucl. Phys. B164 (1980) 171.
\bi{multi}
      I.Ya. Aref'eva, Phys. Lett. B93 (1980) 347;
\\    V.S. Dotsenko and S.N. Vergeles, Nucl. Phys. B169 (1980) 527.
\bi{PT}
      H.T. Nieh and Y.-P. Yao, Phys. Rev. Lett. 32 (1974) 1074;
                             Phys. Rev. D13 (1976) 1082;
\\    B.M. McCoy and T.T. Wu, Phys. Rev. Lett. 35 (1975) 604;
                            Phys. Rev. D12 (1975) 3257;
\\    L. Tyburski, Phys. Rev. D13 (1976) 1107;
\\    C.Y. Lo and H. Cheng, Phys. Rev. D13 (1976) 1131.
\bi{DY}
      J.C. Collins and D. Soper, Nucl. Phys. B197 (1982) 446.
\bi{cusp}
      G.P. Korchemsky and A.V. Radyushkin, Nucl. Phys. B283 ( 1987) 342.
\eb

\newpage

\figures

\begin{center}
\unitlength=0.50mm
\linethickness{0.4pt}
\begin{picture}(169.00,43.00)
\put(10.00,10.00){\vector(1,1){30.00}}
\put(135.00,10.00){\line(1,1){14.92}}
\put(150.00,25.00){\vector(-1,1){15.02}}
\put(166.00,10.00){\line(-1,1){15.09}}
\put(151.00,25.00){\vector(1,1){15.05}}
\put(7.00,43.00){\makebox(0,0)[cc]{$i'$}}
\put(43.00,43.00){\makebox(0,0)[cc]{$j'$}}
\put(132.00,43.00){\makebox(0,0)[cc]{$i'$}}
\put(7.00,7.00){\makebox(0,0)[cc]{$j$}}
\put(43.00,7.00){\makebox(0,0)[cc]{$i$}}
\put(132.00,7.00){\makebox(0,0)[cc]{$j$}}
\put(169.00,7.00){\makebox(0,0)[cc]{$i$}}
\put(169.00,43.00){\makebox(0,0)[cc]{$j'$}}
\put(25.00,2.00){\makebox(0,0)[cc]{$(a)$}}
\put(151.00,2.00){\makebox(0,0)[cc]{$(b)$}}
\put(24.00,26.00){\vector(-1,1){13.96}}
\put(40.00,10.00){\line(-1,1){13.96}}
\put(-15,25){\makebox(0,0)[cc]{$(W_1)^{i'j'}_{ij}=$}}
\put(110,25){\makebox(0,0)[cc]{$(W_2)^{i'j'}_{ij}=$}}
\end{picture}

\bigskip

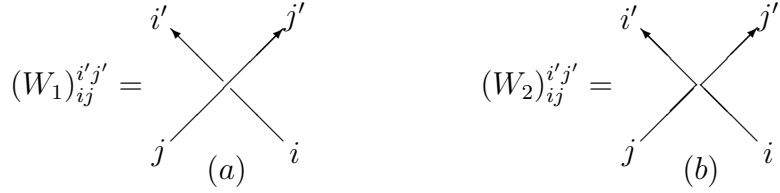
Fig.1: Integration paths (a) and (b) entering into the definition of
the Wilson lines $W_1$ and $W_2$, respectively.
\end{center}

\vspace*{10mm}

\begin{center}
\unitlength=0.50mm
\linethickness{0.4pt}
\begin{picture}(190.00,41.00)
\put(10.00,10.00){\vector(1,1){30.00}}
\put(25.00,2.00){\makebox(0,0)[cc]{$(1)$}}
\put(24.00,26.00){\vector(-1,1){13.96}}
\put(40.00,10.00){\line(-1,1){13.96}}
\put(60.00,10.00){\vector(1,1){30.00}}
\put(75.00,2.00){\makebox(0,0)[cc]{$(2)$}}
\put(74.00,26.00){\vector(-1,1){13.96}}
\put(90.00,10.00){\line(-1,1){13.96}}
\put(110.00,10.00){\vector(1,1){30.00}}
\put(125.00,2.00){\makebox(0,0)[cc]{$(3)$}}
\put(124.00,26.00){\vector(-1,1){13.96}}
\put(140.00,10.00){\line(-1,1){13.96}}
\put(160.00,10.00){\vector(1,1){30.00}}
\put(175.00,2.00){\makebox(0,0)[cc]{$(4)$}}
\put(174.00,26.00){\vector(-1,1){13.96}}
\put(190.00,10.00){\line(-1,1){13.96}}
\bezier{8}(17.00,17.00)(10.00,25.00)(17.00,33.00)
\bezier{8}(67.00,33.00)(75.00,40.00)(83.00,33.00)
\bezier{16}(118.00,32.00)(138.00,38.00)(132.00,18.00)
\bezier{10}(164.00,36.00)(176.00,41.00)(172.00,28.00)
\end{picture}

\bigskip

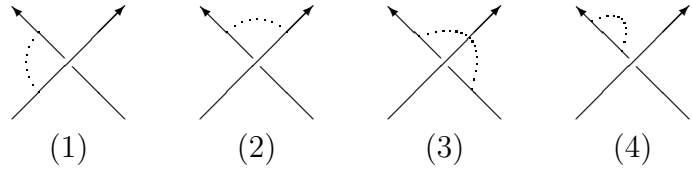
Fig.2: One-loop diagrams contributing to the line function $W_1$.
Solid line represents the integration path, dotted lines denote
gluons.
\end{center}


\bigskip

\begin{center}
\unitlength=0.50mm
\linethickness{0.4pt}
\begin{picture}(282.05,202.00)
\put(10.00,120.00){\line(1,1){14.92}}
\put(25.00,135.00){\vector(-1,1){15.02}}
\put(41.00,120.00){\line(-1,1){15.09}}
\put(26.00,135.00){\vector(1,1){15.05}}
\put(26.00,112.00){\makebox(0,0)[cc]{$(8)$}}
\put(50.00,120.00){\line(1,1){14.92}}
\put(65.00,135.00){\vector(-1,1){15.02}}
\put(81.00,120.00){\line(-1,1){15.09}}
\put(66.00,135.00){\vector(1,1){15.05}}
\put(66.00,112.00){\makebox(0,0)[cc]{$(9)$}}
\put(90.00,120.00){\line(1,1){14.92}}
\put(105.00,135.00){\vector(-1,1){15.02}}
\put(121.00,120.00){\line(-1,1){15.09}}
\put(106.00,135.00){\vector(1,1){15.05}}
\put(106.00,112.00){\makebox(0,0)[cc]{$(10)$}}
\put(10.00,70.00){\line(1,1){14.92}}
\put(25.00,85.00){\vector(-1,1){15.02}}
\put(41.00,70.00){\line(-1,1){15.09}}
\put(26.00,85.00){\vector(1,1){15.05}}
\put(26.00,62.00){\makebox(0,0)[cc]{$(15)$}}
\put(50.00,70.00){\line(1,1){14.92}}
\put(65.00,85.00){\vector(-1,1){15.02}}
\put(81.00,70.00){\line(-1,1){15.09}}
\put(66.00,85.00){\vector(1,1){15.05}}
\put(66.00,62.00){\makebox(0,0)[cc]{$(16)$}}
\put(90.00,70.00){\line(1,1){14.92}}
\put(105.00,85.00){\vector(-1,1){15.02}}
\put(121.00,70.00){\line(-1,1){15.09}}
\put(106.00,85.00){\vector(1,1){15.05}}
\put(106.00,62.00){\makebox(0,0)[cc]{$(17)$}}
\put(10.00,20.00){\line(1,1){14.92}}
\put(25.00,35.00){\vector(-1,1){15.02}}
\put(41.00,20.00){\line(-1,1){15.09}}
\put(26.00,35.00){\vector(1,1){15.05}}
\put(26.00,12.00){\makebox(0,0)[cc]{$(22)$}}
\put(50.00,20.00){\line(1,1){14.92}}
\put(65.00,35.00){\vector(-1,1){15.02}}
\put(81.00,20.00){\line(-1,1){15.09}}
\put(66.00,35.00){\vector(1,1){15.05}}
\put(66.00,12.00){\makebox(0,0)[cc]{$(23)$}}
\put(90.00,20.00){\line(1,1){14.92}}
\put(105.00,35.00){\vector(-1,1){15.02}}
\put(121.00,20.00){\line(-1,1){15.09}}
\put(106.00,35.00){\vector(1,1){15.05}}
\put(106.00,12.00){\makebox(0,0)[cc]{$(24)$}}
\put(10.00,170.00){\line(1,1){14.92}}
\put(25.00,185.00){\vector(-1,1){15.02}}
\put(41.00,170.00){\line(-1,1){15.09}}
\put(26.00,185.00){\vector(1,1){15.05}}
\put(26.00,162.00){\makebox(0,0)[cc]{$(1)$}}
\put(50.00,170.00){\line(1,1){14.92}}
\put(65.00,185.00){\vector(-1,1){15.02}}
\put(81.00,170.00){\line(-1,1){15.09}}
\put(66.00,185.00){\vector(1,1){15.05}}
\put(66.00,162.00){\makebox(0,0)[cc]{$(2)$}}
\put(90.00,170.00){\line(1,1){14.92}}
\put(105.00,185.00){\vector(-1,1){15.02}}
\put(121.00,170.00){\line(-1,1){15.09}}
\put(106.00,185.00){\vector(1,1){15.05}}
\put(106.00,162.00){\makebox(0,0)[cc]{$(3)$}}
\put(131.00,120.00){\line(1,1){14.92}}
\put(146.00,135.00){\vector(-1,1){15.02}}
\put(162.00,120.00){\line(-1,1){15.09}}
\put(147.00,135.00){\vector(1,1){15.05}}
\put(147.00,112.00){\makebox(0,0)[cc]{$(11)$}}
\put(171.00,120.00){\line(1,1){14.92}}
\put(186.00,135.00){\vector(-1,1){15.02}}
\put(202.00,120.00){\line(-1,1){15.09}}
\put(187.00,135.00){\vector(1,1){15.05}}
\put(187.00,112.00){\makebox(0,0)[cc]{$(12)$}}
\put(211.00,120.00){\line(1,1){14.92}}
\put(226.00,135.00){\vector(-1,1){15.02}}
\put(242.00,120.00){\line(-1,1){15.09}}
\put(227.00,135.00){\vector(1,1){15.05}}
\put(227.00,112.00){\makebox(0,0)[cc]{$(13)$}}
\put(131.00,70.00){\line(1,1){14.92}}
\put(146.00,85.00){\vector(-1,1){15.02}}
\put(162.00,70.00){\line(-1,1){15.09}}
\put(147.00,85.00){\vector(1,1){15.05}}
\put(147.00,62.00){\makebox(0,0)[cc]{$(18)$}}
\put(171.00,70.00){\line(1,1){14.92}}
\put(186.00,85.00){\vector(-1,1){15.02}}
\put(202.00,70.00){\line(-1,1){15.09}}
\put(187.00,85.00){\vector(1,1){15.05}}
\put(187.00,62.00){\makebox(0,0)[cc]{$(19)$}}
\put(211.00,70.00){\line(1,1){14.92}}
\put(226.00,85.00){\vector(-1,1){15.02}}
\put(242.00,70.00){\line(-1,1){15.09}}
\put(227.00,85.00){\vector(1,1){15.05}}
\put(227.00,62.00){\makebox(0,0)[cc]{$(20)$}}
\put(131.00,20.00){\line(1,1){14.92}}
\put(146.00,35.00){\vector(-1,1){15.02}}
\put(162.00,20.00){\line(-1,1){15.09}}
\put(147.00,35.00){\vector(1,1){15.05}}
\put(147.00,12.00){\makebox(0,0)[cc]{$(25)$}}
\put(171.00,20.00){\line(1,1){14.92}}
\put(186.00,35.00){\vector(-1,1){15.02}}
\put(202.00,20.00){\line(-1,1){15.09}}
\put(187.00,35.00){\vector(1,1){15.05}}
\put(187.00,12.00){\makebox(0,0)[cc]{$(26)$}}
\put(211.00,20.00){\line(1,1){14.92}}
\put(226.00,35.00){\vector(-1,1){15.02}}
\put(242.00,20.00){\line(-1,1){15.09}}
\put(227.00,35.00){\vector(1,1){15.05}}
\put(227.00,12.00){\makebox(0,0)[cc]{$(27)$}}
\put(131.00,170.00){\line(1,1){14.92}}
\put(146.00,185.00){\vector(-1,1){15.02}}
\put(162.00,170.00){\line(-1,1){15.09}}
\put(147.00,185.00){\vector(1,1){15.05}}
\put(147.00,162.00){\makebox(0,0)[cc]{$(4)$}}
\put(171.00,170.00){\line(1,1){14.92}}
\put(186.00,185.00){\vector(-1,1){15.02}}
\put(202.00,170.00){\line(-1,1){15.09}}
\put(187.00,185.00){\vector(1,1){15.05}}
\put(187.00,162.00){\makebox(0,0)[cc]{$(5)$}}
\put(211.00,170.00){\line(1,1){14.92}}
\put(226.00,185.00){\vector(-1,1){15.02}}
\put(242.00,170.00){\line(-1,1){15.09}}
\put(227.00,185.00){\vector(1,1){15.05}}
\put(227.00,162.00){\makebox(0,0)[cc]{$(6)$}}
\put(251.00,120.00){\line(1,1){14.92}}
\put(266.00,135.00){\vector(-1,1){15.02}}
\put(282.00,120.00){\line(-1,1){15.09}}
\put(267.00,135.00){\vector(1,1){15.05}}
\put(267.00,112.00){\makebox(0,0)[cc]{$(14)$}}
\put(251.00,70.00){\line(1,1){14.92}}
\put(266.00,85.00){\vector(-1,1){15.02}}
\put(282.00,70.00){\line(-1,1){15.09}}
\put(267.00,85.00){\vector(1,1){15.05}}
\put(267.00,62.00){\makebox(0,0)[cc]{$(21)$}}
\put(251.00,20.00){\line(1,1){14.92}}
\put(266.00,35.00){\vector(-1,1){15.02}}
\put(282.00,20.00){\line(-1,1){15.09}}
\put(267.00,35.00){\vector(1,1){15.05}}
\put(267.00,12.00){\makebox(0,0)[cc]{$(28)$}}
\put(251.00,170.00){\line(1,1){14.92}}
\put(266.00,185.00){\vector(-1,1){15.02}}
\put(282.00,170.00){\line(-1,1){15.09}}
\put(267.00,185.00){\vector(1,1){15.05}}
\put(267.00,162.00){\makebox(0,0)[cc]{$(7)$}}
\bezier{8}(16.00,176.00)(10.00,185.00)(16.00,194.00)
\bezier{6}(20.00,190.00)(25.00,196.00)(31.00,190.00)
\bezier{7}(59.00,191.00)(52.00,185.00)(59.00,179.00)
\bezier{21}(55.00,195.00)(81.00,201.00)(76.00,175.00)
\bezier{9}(100.00,180.00)(90.00,185.00)(94.00,196.00)
\bezier{9}(100.00,190.00)(90.00,185.00)(94.00,174.00)
\bezier{8}(138.00,193.00)(131.00,185.00)(138.00,177.00)
\bezier{11}(134.00,197.00)(147.00,202.00)(142.00,189.00)
\bezier{6}(181.00,180.00)(175.00,185.00)(181.00,190.00)
\bezier{11}(176.00,195.00)(166.00,185.00)(176.00,175.00)
\put(218.00,193.00){\circle*{5.20}}
\bezier{12}(256.00,195.00)(245.00,185.00)(256.00,175.00)
\bezier{8}(258.00,193.00)(268.00,197.00)(264.00,187.00)
\bezier{8}(16.00,144.00)(10.00,135.00)(16.00,126.00)
\bezier{10}(21.00,139.00)(33.00,144.00)(30.00,131.00)
\bezier{8}(55.00,145.00)(66.00,149.00)(71.00,140.00)
\bezier{8}(60.00,140.00)(66.00,149.00)(76.00,145.00)
\bezier{8}(97.00,143.00)(105.00,150.00)(114.00,143.00)
\bezier{10}(93.00,147.00)(88.00,136.00)(101.00,139.00)
\bezier{6}(141.00,140.00)(146.00,145.00)(152.00,140.00)
\bezier{10}(136.00,145.00)(146.00,153.00)(157.00,145.00)
\bezier{10}(176.00,145.00)(186.00,152.00)(197.00,145.00)
\bezier{7}(179.00,142.00)(176.00,133.00)(184.00,137.00)
\bezier{10}(216.00,145.00)(226.00,152.00)(237.00,145.00)
\bezier{7}(220.00,141.00)(214.00,135.00)(220.00,129.00)
\bezier{10}(256.00,145.00)(266.00,153.00)(277.00,145.00)
\bezier{13}(261.00,140.00)(256.00,125.00)(272.00,130.00)
\bezier{5}(21.00,89.00)(25.00,93.00)(30.00,89.00)
\bezier{19}(16.00,94.00)(10.00,71.00)(35.00,76.00)
\bezier{14}(56.00,94.00)(52.00,76.00)(69.00,82.00)
\bezier{13}(62.00,88.00)(79.00,93.00)(74.00,77.00)
\bezier{18}(97.00,93.00)(119.00,100.00)(114.00,77.00)
\bezier{10}(93.00,97.00)(89.00,84.00)(101.00,89.00)
\bezier{12}(141.00,90.00)(137.00,75.00)(151.00,81.00)
\bezier{21}(136.00,95.00)(162.00,102.00)(156.00,76.00)
\bezier{21}(175.00,96.00)(202.00,101.00)(198.00,75.00)
\bezier{7}(179.00,92.00)(175.00,84.00)(184.00,87.00)
\bezier{8}(218.00,93.00)(226.00,100.00)(235.00,93.00)
\bezier{8}(218.00,77.00)(227.00,70.00)(235.00,77.00)
\bezier{8}(258.00,93.00)(251.00,85.00)(258.00,77.00)
\bezier{8}(275.00,93.00)(282.00,85.00)(275.00,77.00)
\bezier{7}(19.00,41.00)(13.00,35.00)(19.00,29.00)
\bezier{8}(17.00,43.00)(21.00,52.00)(12.00,48.00)
\bezier{7}(59.00,41.00)(53.00,35.00)(59.00,29.00)
\bezier{7}(74.00,43.00)(70.00,51.00)(79.00,48.00)
\bezier{6}(99.00,41.00)(105.00,46.00)(112.00,41.00)
\bezier{8}(92.00,48.00)(101.00,52.00)(97.00,43.00)
\bezier{7}(140.00,41.00)(146.00,47.00)(153.00,41.00)
\bezier{8}(138.00,27.00)(129.00,31.00)(133.00,22.00)
\bezier{8}(178.00,27.00)(169.00,31.00)(173.00,22.00)
\bezier{13}(181.00,30.00)(176.00,46.00)(192.00,40.00)
\bezier{14}(221.00,30.00)(215.00,46.00)(232.00,40.00)
\bezier{8}(235.00,27.00)(244.00,32.00)(240.00,22.00)
\bezier{14}(258.00,27.00)(252.00,44.00)(270.00,39.00)
\bezier{18}(258.00,44.00)(280.00,50.00)(274.00,28.00)
\end{picture}

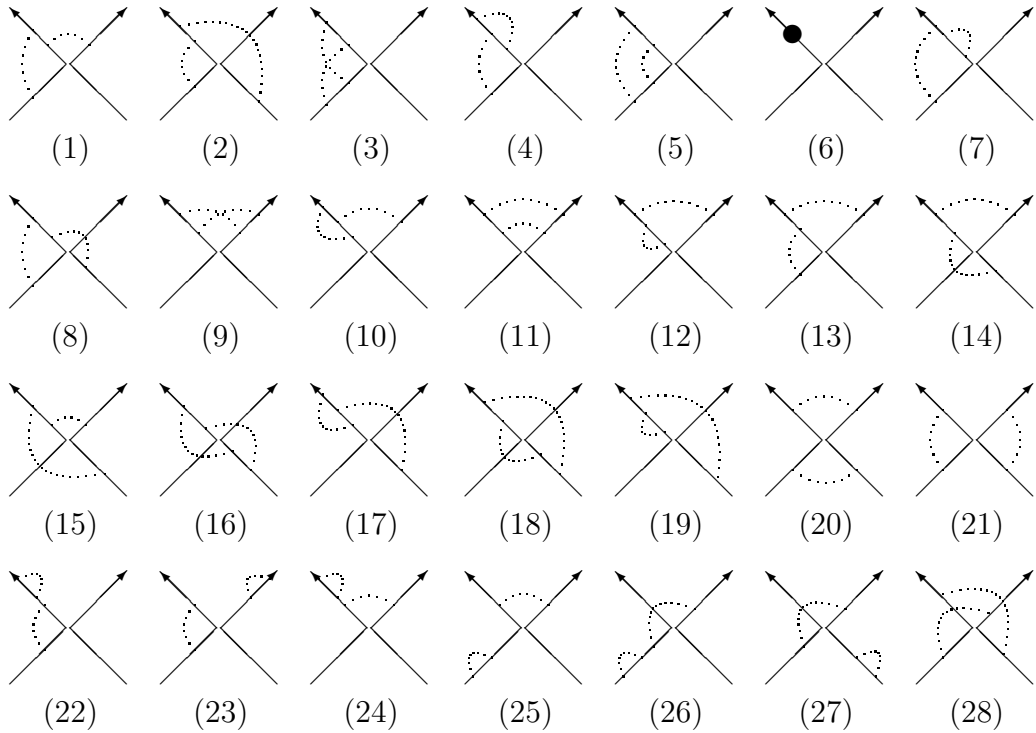
Fig.3: Two-loop QED like diagrams contributing to the line function $W_2$.
The blob in the diagram (6) denotes all possible QED like ``self-energy''
corrections to the Wilson line.
\end{center}

\bigskip

\begin{center}
\unitlength=0.50mm
\linethickness{0.4pt}
\begin{picture}(160.05,41.00)
\put(10.00,10.00){\line(1,1){14.92}}
\put(25.00,25.00){\vector(-1,1){15.02}}
\put(41.00,10.00){\line(-1,1){15.09}}
\put(26.00,25.00){\vector(1,1){15.05}}
\put(26.00,2.00){\makebox(0,0)[cc]{$(1)$}}
\put(49.00,10.00){\line(1,1){14.92}}
\put(64.00,25.00){\vector(-1,1){15.02}}
\put(80.00,10.00){\line(-1,1){15.09}}
\put(65.00,25.00){\vector(1,1){15.05}}
\put(65.00,2.00){\makebox(0,0)[cc]{$(2)$}}
\put(90.00,10.00){\line(1,1){14.92}}
\put(105.00,25.00){\vector(-1,1){15.02}}
\put(121.00,10.00){\line(-1,1){15.09}}
\put(106.00,25.00){\vector(1,1){15.05}}
\put(106.00,2.00){\makebox(0,0)[cc]{$(3)$}}
\put(129.00,10.00){\line(1,1){14.92}}
\put(144.00,25.00){\vector(-1,1){15.02}}
\put(160.00,10.00){\line(-1,1){15.09}}
\put(145.00,25.00){\vector(1,1){15.05}}
\put(145.00,2.00){\makebox(0,0)[cc]{$(4)$}}
\bezier{9}(16.00,34.00)(9.00,25.00)(16.00,16.00)
\bezier{10}(53.00,36.00)(64.00,41.00)(60.00,29.00)
\bezier{9}(96.00,34.00)(105.00,41.00)(115.00,34.00)
\bezier{16}(150.00,20.00)(153.00,40.00)(134.00,35.00)
\put(60.00,37.00){\circle*{4.00}}
\put(106.00,38.00){\circle*{4.00}}
\put(147.00,34.00){\circle*{4.00}}
\put(12.00,25.00){\circle*{4.47}}
\end{picture}

\bigskip

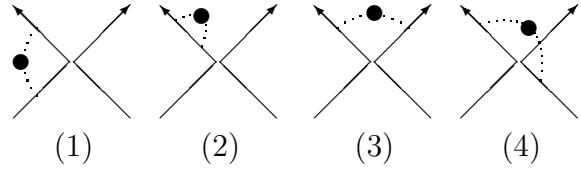
Fig.4: Two-loop self-energy diagrams. The blob denotes one-loop
self-energy corrections to the gluon propagator.
\end{center}

\bigskip

\begin{center}
\unitlength=0.50mm
\linethickness{0.4pt}
\begin{picture}(121.05,41.00)
\put(10.00,10.00){\line(1,1){14.92}}
\put(25.00,25.00){\vector(-1,1){15.02}}
\put(41.00,10.00){\line(-1,1){15.09}}
\put(26.00,25.00){\vector(1,1){15.05}}
\put(26.00,2.00){\makebox(0,0)[cc]{$(1)$}}
\put(49.00,10.00){\line(1,1){14.92}}
\put(64.00,25.00){\vector(-1,1){15.02}}
\put(80.00,10.00){\line(-1,1){15.09}}
\put(65.00,25.00){\vector(1,1){15.05}}
\put(65.00,2.00){\makebox(0,0)[cc]{$(2)$}}
\put(90.00,10.00){\line(1,1){14.92}}
\put(105.00,25.00){\vector(-1,1){15.02}}
\put(121.00,10.00){\line(-1,1){15.09}}
\put(106.00,25.00){\vector(1,1){15.05}}
\put(106.00,2.00){\makebox(0,0)[cc]{$(3)$}}
\bezier{9}(35.00,34.00)(42.00,24.00)(37.00,14.00)
\bezier{4}(39.00,26.00)(35.00,24.00)(31.00,20.00)
\bezier{10}(53.00,36.00)(66.00,41.00)(74.00,34.00)
\bezier{5}(66.00,38.00)(64.00,32.00)(59.00,30.00)
\bezier{9}(96.00,34.00)(105.00,40.00)(117.00,36.00)
\bezier{8}(107.00,38.00)(112.00,29.00)(112.00,19.00)
\end{picture}

\bigskip

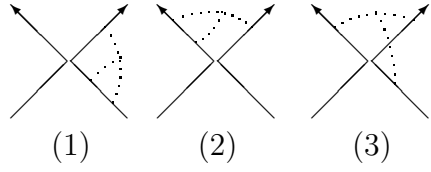
Fig.5: Two-loop three gluon diagrams.
\end{center}

\end{document}